\documentstyle[12pt]{article}

\font\blackboard=msbm10 at 12pt
\font\blackboards=msbm7
\font\blackboardss=msbm5
\newfam\black
\textfont\black=\blackboard
\scriptfont\black=\blackboards
\scriptscriptfont\black=\blackboardss
\def\bb#1{{\fam\black\relax#1}}
\def\BZ{\bb Z}
\def\br{\bb R}
\def\bff{{\bf F}}
\def\bfx{{\bf X}}
\def\bfy{{\bf Y}}

\def\Re{{\rm Re\ }}
\def\Im{{\rm Im\ }}

\newcommand{\junk}[1]{}

\newcommand{\ba}{\begin{array}}
\newcommand{\ea}{\end{array}}
\newcommand{\be}{\begin{equation}}
\newcommand{\ee}{\end{equation}}
\newcommand{\bea}{\begin{eqnarray}}
\newcommand{\eea}{\end{eqnarray}}
\newcommand{\beas}{\begin{eqnarray*}}
\newcommand{\eeas}{\end{eqnarray*}}

\font\cmss = cmss12
\def\half{{1 \over 2}}
\def\identity{{\rlap{1} \hskip 1.6pt \hbox{1}}}
\def\integer{{\rlap{\cmss Z} \hskip 1.8pt \hbox{\cmss Z}}}
\def\laplace{{\kern1pt\vbox{\hrule height 1.2pt\hbox{\vrule width
1.2pt\hskip
  3pt\vbox{\vskip 6pt}\hskip 3pt\vrule width 0.6pt}\hrule height
  0.6pt}
  \kern1pt}}
\def\scriptlap{{\kern1pt\vbox{\hrule height 0.8pt\hbox{\vrule width
  0.8pt
  \hskip2pt\vbox{\vskip 4pt}\hskip 2pt\vrule width 0.4pt}\hrule height
  0.4pt}
  \kern1pt}}
\def\slash#1{{\rlap{$#1$} \thinspace /}}
\def\roughly#1{\raise.3ex\hbox{$#1$\kern-
.75em\lower1ex\hbox{$\sim$}}}
 
\def\complex{{\hbox{\cmss C} \llap{\vrule height 7.0pt
width 0.4pt depth -.4pt \hskip 0.5 pt \phantom .}}}
\def\real{{\hbox{\cmss R} \llap{\vrule height 7.1pt width 0.4pt
depth -.1pt \hskip 0.6 pt \phantom .}}}
\def\projectiveB{{\hbox{\cmss P} \llap{\vrule height 7.1pt width 0.4pt
depth -.1pt \hskip 0.6 pt \phantom .}}}

\def\I{{\hbox{\cmss I}}}
\def\IA{\mbox{{\hbox{\cmss IA}}}}
\def\IB{\mbox{{\hbox{\cmss IB}}}}
\def\II{\mbox{{\rlap{\cmss I} \hskip 0.7 true pt \hbox{\cmss I}}}}
\def\IIA{\mbox{{\rlap{\cmss I} \hskip 0.7 true pt \hbox{\cmss IA}}}}
\def\IIB{\mbox{{\rlap{\cmss I} \hskip 0.7 true pt \hbox{\cmss IB}}}}

\def\gym{g^2_{\scriptscriptstyle YM}}

\def\str{{\rm STr} \,}
\def\sym{{\rm Sym} \,}

\def\tr{{\rm Tr} \,}
\def\trl{{\rm Tr}_L}
\def\trg{{\rm Tr}_G}

\def\ab{{\bar{\alpha}}}
\def\bb{{\bar{\beta}}}
\def\bm{{\bar{\mu}}}
\def\bn{{\bar{\nu}}}
\def\ha{{\hat{a}}}
\def\hb{{\hat{b}}}

\newcommand{\NP}{{\em Nucl.\ Phys.\ }}
\newcommand{\AP}{{\em Ann.\ Phys.\ }}
\newcommand{\PL}{{\em Phys.\ Lett.\ }}
\newcommand{\PR}{{\em Phys.\ Rev.\ }}
\newcommand{\PRP}{{\em Phys.\ Rep.\ }}
\newcommand{\CMP}{{\em Comm.\ Math.\ Phys.\ }}
\newcommand{\MPL}{{\em Mod.\ Phys.\ Lett.\ }}
\newcommand{\PRL}{{\em Phys.\ Rev.\ Lett.\ }}
\newcommand{\IJMP}{{\em Int.\ J.\ Mod.\ Phys.\ }}

\newcommand{\um}[1]{\"{#1}}
\newcommand{\re}{{\rm Re}}
\newcommand{\im}{{\rm Im}}
\newcommand{\lag}{{\cal L}}
\newcommand{\newcaption}[1]{\centerline{\parbox{6in}{\caption{#1}}}}
\newcommand{\fslash}{F\!\!\!\!/\ }
\newcommand{\inn}{\!\cdot\!}

\newcommand{\gone}[1]{}
\begin{document}
\pagestyle{plain}
\setcounter{page}{1}

\baselineskip16pt

\begin{titlepage}

\begin{flushright}
MIT-CTP-2954\\
hep-th/0002237
\end{flushright}
\vspace{13 mm}

\begin{center}

{\Large \bf Level truncation and the tachyon\\[0.3cm] in open bosonic string
field theory}
%\vspace{3mm}

\end{center}

\vspace{7 mm}

\begin{center}

Nicolas Moeller and Washington Taylor

\vspace{3mm}
{\small \sl Center for Theoretical Physics} \\
{\small \sl MIT, Bldg.  6-306} \\
{\small \sl Cambridge, MA 02139, U.S.A.} \\
{\small \tt moeller@pierre.mit.edu, wati@mit.edu}\\
\end{center}

\vspace{8 mm}

\begin{abstract}
The tachyonic instability of the open bosonic string is analyzed using
the level truncation approach to string field theory.  We have
calculated all terms in the cubic action of the string field theory
describing zero-momentum interactions of up to level 20 between
scalars of level 10 or less.  These results are used to study the
tachyon effective potential and the nonperturbative stable vacuum.  We
find that the energy gap between the unstable and stable vacua
converges much more quickly than the coefficients of the effective
tachyon potential.  By including fields up to level 10, 99.91\% of the
energy from the bosonic D-brane tension is cancelled in the
nonperturbative stable vacuum.  It appears that the perturbative
expansion of the effective tachyon potential around the unstable
vacuum has a small but finite radius of convergence.  We find evidence
for a critical point in the tachyon effective potential at a small
negative value of the tachyon field corresponding to this radius of
convergence.  We study the branch structure of the effective potential
in the vicinity of this point and speculate that the tachyon effective
potential is globally nonnegative.
\end{abstract}

%\vspace{2cm}
\vspace{1cm}
\begin{flushleft}
February 2000
\end{flushleft}
\end{titlepage}
\newpage

\section{Introduction}

The appearance of a tachyon in the spectrum of both open and closed
bosonic strings has appeared to be a fundamental obstacle to a
physical interpretation of these theories since the early days of the
dual resonance model.  Early work on the subject indicated the
possible existence of a more stable nonperturbative vacuum which can
be reached by a condensation of the tachyon and other string fields in
the open bosonic string theory \cite{Bardakci-tachyon,ks-open}.  At
that time, however, the connection between open strings and Dirichlet
branes \cite{Polchinski} had not yet been realized, so that the
significance of the nonperturbative stable vacuum was not widely
appreciated.

It was recently pointed out by Sen \cite{Sen-universality} that the
condensation of the tachyon in open bosonic string theory should
correspond to the process of annihilation of an unstable D25-brane.
The nonperturbative stable vacuum of the open string should simply be
the vacuum corresponding to empty space, and the energy difference
between the stable and unstable vacua should therefore be given by
the mass of the D25-brane.  Sen suggested that it should be
possible to precisely calculate this energy gap using open string
field theory.  In fact, it was found over a decade ago by Kostelecky
and Samuel that truncating open bosonic string field theory at low
mass levels gives a systematic approximation scheme which seems to
converge to a finite value for the energy difference between the
unstable and stable vacua \cite{ks-open,Kostelecky-Potting}.  Sen and
Zwiebach have carried out a calculation of this type and have shown
that including fields up to mass level 4 and interactions up to mass
level 8 gives a mass gap of 98.6\% of the D25-brane
energy.   Similar calculations of tachyon condensation have
recently been performed in open superstring field theory
\cite{Berkovits-tachyon,bsz}.  The level-truncation approach to string
field theory has also been used to study lower-dimensional D$p$-branes
as solitonic lumps \cite{Harvey-Kraus,bsz}.  In all these
calculations, truncation of string field theory to the first few
levels seems to give a sequence of successively better approximations
to nonperturbative physical quantities.

In this paper we extend the level truncation approach to open bosonic
string field theory to include scalar fields of mass levels $> 4$,
following earlier work in \cite{WT-SFT}.  We use the level-truncated
theory as a tool for exploring various features of the tachyon and its
condensation into a stable vacuum.  We compute all terms in the string
field theory action involving fields of level less than or equal to 10
and interactions of up to mass level 20.  We perform a nonperturbative
calculation of the energy gap between the stable and unstable vacua
and find that 99.91\% of the D25-brane tension is produced in the
level (10, 20) truncated theory.

We study the structure of the tachyon effective potential in some
detail using the various level-truncated theories.  We compute the
first 60 coefficients $c_n$ of $\phi^n$ in the effective tachyon
potential in several successive level truncations up to level (10,
20).  We find that the effective tachyon potential has a radius of
convergence which decreases to an apparently finite asymptotic value
as the level number increases.  This radius of convergence is
substantially smaller than the tachyon value corresponding to the
stable vacuum, although the associated singularity is at a negative
value of $\phi$ while the stable vacuum arises at $\phi > 0$.  We
investigate the branch structure of the tachyon effective potential
near the singular point, and use our numerical results to speculate
about the behavior of the effective potential beyond this point.

In Section \ref{sec:truncation} we review the level truncation
approach to string field theory and outline our calculation of the
action up to level (10, 20).  In section \ref{sec:perturbative} we use
our results to analyze the perturbative expansion of the tachyon
effective potential, and in section \ref{sec:nonperturbative} we
discuss the nonperturbative stable vacuum and the branch structure of
the effective tachyon potential.

\section{Level truncation of string field theory}
\label{sec:truncation}

In this section we briefly review the level truncation approach to
string field theory and describe  our calculation of interactions
between scalar string fields of level $\leq 10$.  Throughout this
paper we follow the conventions and notation of \cite{ks-open,WT-SFT}.
For more detailed reviews of open string field theory see
\cite{lpp,Gaberdiel-Zwiebach}.

\subsection{The scalar potential in open bosonic string field theory}

String field
theory is described in terms of a string field $\Phi$ which contains a
component field for every state in the first-quantized string Fock
space.   For the open bosonic string, a particularly simple string
field theory was suggested by Witten \cite{Witten-SFT} in which the
action takes the cubic form
\begin{equation}
S = \frac{1}{2 \alpha'}  \int \Phi \star Q \Phi + \frac{g}{3!}  \int \Phi \star
\Phi \star \Phi,
\label{eq:SFT-action}
%\nonumber
\end{equation}
where $Q$ is the BRST operator and $\star$ is the string field theory star
product.

In Feynman-Siegel gauge, the string field can be expanded in the form
\begin{equation}
\Phi = \left( \phi + A_\mu \alpha^\mu_{-1}
 + \frac{1}{\sqrt{2}} B_{\mu \nu} \alpha^\mu_{-1} \alpha^\nu_{-1} +
 \beta b_{-1} c_{-1}  + \cdots \right)| 0 \rangle
\label{eq:expansion}
\end{equation}
where $| 0 \rangle = c_1 | \Omega \rangle$ is the state in the string
Hilbert space associated with the tachyon field.  The expansion
(\ref{eq:expansion}) contains an infinite series of fields.  The level
of each field in the expansion is defined to be the sum of the level
numbers of the creation operators which act on $| 0 \rangle$ to
produce the associated state.  Thus, the tachyon is the unique field
of level 0, the gauge field $A_\mu$ is the unique field of level 1,
etc.  The states associated with these fields are all in the subspace
${\cal H}$ of the full string Hilbert space containing states of ghost
number 1.

In this paper we are interested in the behavior of the tachyon field
$\phi$.  In particular, we wish to study questions related to the
appearance of a Lorentz-invariant stable vacuum in the theory when the
tachyon and other scalar fields acquire nonzero condensates.  Because
all the questions we will address here involve Lorentz-invariant
phenomena, we can restrict attention to scalar fields in the string
field expansion.  We write the string field expansion in terms of
scalar fields as
\begin{equation}
 \Phi = \sum_{i = 1}^{\infty}  \psi^i | s_i \rangle
\label{eq:expand-scalars}
\end{equation}
where $| s_i \rangle$ are all the scalar states in ${\cal H}$.  The
scalar states at levels 0 and 2 are
\begin{eqnarray*}
| s_1  \rangle & = &  | 0 \rangle\\
| s_2  \rangle & = &  \alpha_{-1}  \cdot\alpha_{-1} | 0 \rangle\\
| s_3  \rangle & = &   b_{-1} c_{-1}| 0 \rangle
\end{eqnarray*}
The associated scalar fields can be related to the fields in the
expansion (\ref{eq:expansion}) through
\begin{eqnarray*}
\phi & = &  \psi^1\\
B_{\mu \nu} & = &  \sqrt{2} \eta_{\mu \nu}  \psi^2\\
\beta & = & \psi^3
\end{eqnarray*}
We will often refer to the tachyon field $\psi^1$ simply as $\phi$.
(Note that a superscript on $\psi$ will always indicate a field index while a
superscript on $\phi$ indicates an exponent.)
For the calculations of interest here, contributions from scalar
fields of odd level cancel due to a twist symmetry
\cite{ks-open,Sen-Zwiebach}.  Thus, we need only consider scalar
fields of even level number.  All scalar states at levels 0, 2, 4 and
6 are listed in appendix A.

An explicit algorithm for computing the terms in the string field
theory action (\ref{eq:SFT-action}) using the oscillator modes of the
matter fields and ghost fields of the bosonic conformal field theory
was given in \cite{Gross-Jevicki-12,cst,Samuel}.  The aspects of this
formalism needed for the computations in this paper are reviewed in
\cite{WT-SFT}.  
The
quadratic term evaluated on a state $| s \rangle$ is simply
\begin{equation}
\int | s \rangle \star Q | s \rangle =
\langle s | c_0 \left(\alpha' p^2 +\frac{1}{2} M^2 \right)| s \rangle
\label{eq:quadratic}
\end{equation}
where $\frac{1}{2}M^2$ is the level
of $s$ minus 1, $p$ is the momentum of the state $s$, $\langle s |$ is
the BPZ dual state to $| s \rangle$ produced by acting with the
conformal transformation $z \rightarrow -1/z$, and the dual
vacuum satisfies $\langle 0 | c_0 | 0 \rangle = 1$.
The cubic interaction terms in the string field action can be
described in terms of Witten's vertex operator $V$ through
\begin{equation}
\int \Phi \star\Phi\star\Phi =
\langle V | (\Phi \otimes \Phi \otimes \Phi)
\label{eq:cubic}
\end{equation}
where $V$ is a state in the tensor product space ${\cal H}^*\otimes
{\cal H}^*\otimes {\cal H}^*$, given in terms of oscillator modes by
\begin{equation}
\langle V | = 
\delta (p_{(1)} +p_{(2)} + p_{(3)})
 (\langle 0 | c_0^{(1)} \otimes \langle 0 |c_0^{(2)}
 \otimes\langle 0 | c_0^{(3)})
\exp \left(\frac{1}{2} \alpha^{(r) \mu}_n N^{rs}_{nm} \eta_{\mu \nu} 
 \alpha^{(s)\nu}_{m} + c^{(r)}_n X^{rs}_{nm} b^{(s)}_m \right).
\label{eq:vertex}
\end{equation}
The coefficients  $N^{rs}_{nm},X^{rs}_{nm}$ can be calculated from
formulae in \cite{Gross-Jevicki-12,cst,Samuel}; explicit tables of
these coefficients for $n, m \leq 8$ are given in \cite{WT-SFT}.

Using the equations (\ref{eq:quadratic}, \ref{eq:cubic}) for the
quadratic and cubic terms in the string field theory action, the
potential for the zero momentum scalar fields in ${\cal H}$ can be written as
\begin{equation}
V = \sum_{i, j} d_{ij} \psi^i \psi^j +
g \kappa \sum_{i, j, k} t_{ijk} \psi^i \psi^j \psi^k
\label{eq:potential}
\end{equation}
where $g$ is the string coupling constant and
\[
\kappa = \frac{3^{7/2}}{2^7} \,.
\]
has been chosen so that $t_{111} = 1$.
Throughout the paper we set $\alpha' = 1$.
The coefficients $d_{ij}, t_{ijk}$ can be explicitly computed
for any given values of the indices.  The action can be truncated by
only including fields up to a fixed level $L$ and interactions
including terms whose total level does not exceed another fixed value
$I$.  In \cite{ks-open}, the complete action with $(L, I) = (2, 6)$
was calculated and shown to be
\begin{eqnarray}
V & = &  -\frac{1}{2}\phi^2 + 26  (\psi_2)^2 -\frac{1}{2}(\psi_3)^2
\nonumber\\
 &  &+ \kappa g\left[\phi^3 + \phi^2 \left(-\frac{5 \cdot 26}{9}
 \psi_2-\frac{11}{9}  \psi_3\right) \right. \nonumber\\
& &\hspace*{0.4in}\left.
+
\phi  \left(\frac{4 \cdot 7 \cdot  13 \cdot 83}{3^5} (\psi_2)^2
+ \frac{4 \cdot 5 \cdot 11 \cdot 13}{3^5}\psi_2 \psi_3
+ \frac{19}{3^4} (\psi_3)^2  \right) \right. \nonumber\\
& &\hspace*{0.4in}\left.
 -\frac{2^3 \cdot 3 \cdot 7 \cdot 13 \cdot 41 \cdot 73}{ 3^9}
 (\psi_2)^3
 -\frac{2^2 \cdot 7 \cdot 13 \cdot  11 \cdot 83}{ 3^8}
 (\psi_2)^2\psi_3\right. \nonumber\\
& &\hspace*{0.4in}\left.
 -\frac{2 \cdot 5 \cdot 13 \cdot 19}{ 3^7}
 \psi_2  (\psi_3)^2
 -\frac{1}{ 3^4}
 (\psi_3)^3\right] \nonumber
\end{eqnarray}
(Note that we have lowered indices on the fields $\psi^2, \psi^3$ for
clarity.)  Results using the complete action with $(L, I) = (4, 12)$
were described in \cite{Kostelecky-Potting}.  In \cite{Sen-Zwiebach}
calculations were performed at level $(L, I) = (4, 8)$.

With the help of the symbolic manipulation program {\it Mathematica},
we have calculated all terms in the action up to levels $(L, I) = (10,
20)$.  There are 252 fields at levels $\leq 10$ and 138,202 distinct
cubic interaction terms between fields whose total level is $\leq 20$,
so it is clearly impractical to reproduce the full action here.  It is
worth mentioning, however, some points which help to simplify the
calculation.

One significant simplification arises from the fact that the matter
and ghost fields almost completely decouple in the action.  In fact,
it is clear from (\ref{eq:vertex}) that this decoupling is complete in
the cubic terms, and from (\ref{eq:quadratic}) that the only coupling
in the quadratic terms arises from the appearance of the total level
of the scalar field in question, including the levels of both the
matter and ghost oscillators needed to produce the state.  Because of
this decoupling, we find that it is convenient to decompose the
Hilbert space into the matter and ghost Hilbert spaces
\begin{equation}
{\cal H} ={\cal H}_{\rm mat} \otimes {\cal H}_{\rm gh}
\label{eq:hmg}
\end{equation}
and to
separately enumerate the scalar fields in
the matter and ghost sectors $|\eta_m \rangle \in{\cal H}_m, |\chi_g
\rangle \in{\cal H}_g$.  The first few states in each of these
component Hilbert spaces are
\begin{eqnarray*}
| \eta_1 \rangle = | \chi_1 \rangle & = &  | 0 \rangle\\
| \eta_2 \rangle & = &    \alpha_{-1}  \cdot\alpha_{-1} | 0 \rangle\\
 | \chi_2 \rangle & = &     b_{-1} c_{-1}| 0 \rangle
\end{eqnarray*}
Given the decomposition (\ref{eq:hmg}) of the Hilbert space we can
write each of the scalar fields $| s_i \rangle$ in ${\cal H}$ as a
tensor product
\begin{equation}
| s_i \rangle= | \eta_{m (i)} \rangle \otimes
| \chi_{g (i)} \rangle
\label{eq:tensor-decomposition}
\end{equation}
Thus, for example, $m (1) = 1$ and $g (1) = 1$.  A list of
matter and ghost scalars up to level 6 is given in appendix B, and the
decomposition of the scalars in ${\cal H}$ into  matter and ghost
factors is given in appendix A for the scalars of level $\leq 6$.

{}From the decomposition into matter and ghost components, we can now
write the quadratic and cubic coefficients
(\ref{eq:quadratic},\ref{eq:cubic}) as
\begin{eqnarray}
d_{ij} & = &  \frac{2 ({\rm level} (i) -1)}{
({\rm level} (m (i)) -1) \cdot ({\rm level}
(g (i)) -1)} 
d^{\rm mat}_{ m (i)m (j)}
d^{\rm gh}_{g (i)g (j)} \label{eq:coefficient-product}\\
t_{ijk} & = &  t^{\rm mat}_{ m (i)m (j)m (k)}
t^{\rm gh}_{g (i)g (j)g (k)} \nonumber
\end{eqnarray}
where $d^{\rm mat}_{mn}, d^{\rm gh}_{gh}, t^{\rm mat}_{mn p}$
and $t^{\rm gh}_{ghj}$ are the quadratic and cubic coefficients in
the matter and ghost sectors, and where level$(i) =$ level$(g(i))+$
level$(m (i))$ indicates the level of the state $| s_i \rangle$.
Quadratic and cubic coefficients of the matter and ghost
scalars appearing in scalar fields up to level 6 are tabulated in
Appendix C.  From these coefficients and
(\ref{eq:coefficient-product}) one can reproduce the
entire string field theory action for zero momentum scalars up to
level 6 fields and level 16 interactions.  We have carried out the
calculation of all matter and ghost interactions up to fields of level
10 and interactions of level 20.  While we do not reproduce here the
complete results of this calculation, we will use these results to
study questions of physical interest in the remainder of the paper.

We end this section with a brief discussion of the approach used in
\cite{Sen-Zwiebach}, in which a reduced Hilbert space of
background-independent fields was used.  It was pointed out in
\cite{Sen-universality} that it is possible to truncate the Hilbert
space ${\cal H}$ in a consistent fashion by considering the subspace
${\cal H}_1$ obtained by considering only those states produced by
acting with the ghost fields $b, c$ and the Virasoro generators
$L_{-n}, n \geq 2$ associated with the stress tensor of the matter
fields.  In the stable vacuum of the theory, only scalar fields in
${\cal H}_1$ acquire nonzero expectation values.  At level 4 and
above, the number of scalars in ${\cal H}_1$ is less than the number
of scalars in ${\cal H}$, so that particularly when the level number
becomes very large the effective action for the level-truncated theory
is significantly simplified by restricting attention to the truncated
Hilbert space.  To compare the number of scalars in ${\cal H}$ at a
fixed level $n$, which we denote $h_n$, with the number of scalars in
${\cal H}_1$, which we denote $h^1_{n}$, we can write generating
functions
\begin{eqnarray}
f (x, y) & = & \prod_{p = 2}^{\infty}  \frac{1}{(1-x^p)^{\lfloor
\frac{p}{2} \rfloor}}  \prod_{q = 1}^{ \infty} 
(1 + x^qy) (1 +  \frac{x^q}{y} ) = \sum_{n, m}h_{n, m} x^ny^m\\
f_1 (x, y) & = & \prod_{p = 2}^{\infty}  \frac{1}{(1-x^p)}
  \prod_{q = 1}^{ \infty} 
(1 + x^qy) (1 +  \frac{x^q}{y} ) = \sum_{n, m}h^1_{n, m} x^ny^m
\end{eqnarray}
in terms of which
\begin{eqnarray*}
h_n & = &  h_{n, 0}\\
h^1_n & = &  h^1_{n, 0}
\end{eqnarray*}
The  number of scalars in ${\cal H}$ and ${\cal H}_1$ at each even
level up to $n = 20$ is tabulated in Table~\ref{t:scalars}.
\begin{table}[htp]
\begin{center}
\begin{tabular}{ ||r  || c | c | c | c | c | c | c |c | c | c | c ||}
\hline
\hline
 $n$ & 0 & 2 & 4 & 6 & 8 & 10 & 12 & 14 & 16 & 18 & 20\\
\hline
\hline
 $h_n$ & 1 &2 & 7 & 21 & 60 & 161 & 415 & 1021 & 2432 & 5620 & 12639\\
\hline
 $h^1_n$ & 1 &2 & 6 & 17 & 43 & 102 & 231  & 496 & 1027 & 2060 & 4010 \\
\hline
\hline
\end{tabular}
\end{center}
\caption[x]{\footnotesize The number of scalars $h_n$ and $h^1_n$ in
${\cal H}$ and ${\cal H}_1$ at level $n$}
\label{t:scalars}
\end{table}
Up to the level 10 fields we consider in this paper, the difference
between ${\cal H}$ and ${\cal H}_1$ is less than a factor of 2, so
that there is no extraordinary advantage to be gained by using the
smaller Hilbert space.  Clearly, however, as the level becomes much
larger than 10, the reduced size of the truncated Hilbert space would
make explicit calculations much easier, assuming that calculations in
${\cal H}_1$ could be done just as efficiently as corresponding
calculations in ${\cal H}$.

In our calculations we have worked with the complete oscillator
Hilbert space ${\cal H}$ and not with ${\cal H}_1$.  The main reason
for this is that at this point no method has been developed for
computing cubic interactions between fields in ${\cal H}_1$ which is
as systematic and efficient as the oscillator method described above
for computing cubic interactions between fields in ${\cal H}$.  It is
possible to calculate cubic interactions in ${\cal H}_1$ by computing
the relevant correlation functions in the open bosonic string theory
\cite{lpp}, but this method is somewhat more complicated than the
straightforward oscillator approach.  Of course, the interactions in
${\cal H}_1$ can always be computed by rewriting each state in terms
of the oscillator basis and then using (\ref{eq:vertex}), but this
requires just as much work as the computation directly in the
oscillator basis.  In any case, the results in Appendix C for fields
up to level 6 are given in the oscillator basis of ${\cal H}$; these
interactions can easily be translated into an action for fields in
${\cal H}_1$ by explicitly writing the Virasoro generators $L_{-n}$ in
terms of the oscillator basis.

\section{Tachyon effective potential: perturbation expansion}
\label{sec:perturbative}

Given the  (rational) coefficients  in the scalar potential
(\ref{eq:potential}) truncated at level $(L, I)$, we can perform a number of
interesting calculations.  In particular, we can study the effective
potential of the tachyon field and identify the stable vacuum or vacua
of the theory.

An effective potential for the tachyon field $\phi$ can be determined
by starting with the complete set of terms in the cubic potential
(\ref{eq:potential}) truncated at level $(L, I)$, fixing a value of
$\phi = \psi^1$, solving for all fields $\psi^i, i > 1$, and plugging
back into the potential to rewrite it as a function of $\phi$.  If
there are $N$ scalar fields involved, this means that we need to solve
a system of $N-1$ simultaneous quadratic equations.  In principle
there are many solutions of this system of equations, but if we are
interested in the branch of the solution on which all fields vanish
when $\phi = 0$, it is easy to determine which branch to choose for
each of the quadratic equations (an explicit example of this choice
is described below).  Clearly when $N$ becomes large it is
impractical to find an exact analytic form for the effective tachyon
potential.  Various numerical methods can be used to approximate the
effective potential arising from integrating out a large number of
fields; for example,
a numerical analysis of the effective potential was performed
in \cite{ks-open} using the level 2 action with the two fields
$\psi^2, \psi^3$ integrated
out.  In section \ref{sec:effective} we extend these results by
analyzing the structure of the
effective potential at levels 4 and 6 using numerical methods.

Another approach to studying the effective potential, which we
consider in this section, is to determine the terms in a power series
expansion of the potential around the unstable vacuum $\phi = 0$.  In
subsection \ref{sec:tree-algebra} we describe an algebraic method for
efficiently summing all diagrams contributing to the $\phi^n$ term in
the effective potential.  In subsection \ref{sec:coefficients} we
summarize the results of our calculation of these coefficients up to
$n = 60$ at level $L \leq 10$, and discuss the implications of these results.
In particular, we find that the power series expansion of the tachyon
effective potential seems to have a radius of convergence which
approaches a finite but nonzero value as the level number is
increased.  As we will discuss in the next section, the
stable vacuum of the theory lies outside the radius of convergence of
this power series.

\subsection{Summing planar diagrams}
\label{sec:tree-algebra}

In the vicinity of the unstable vacuum $\psi^i = 0$ we can
perform a power series expansion of the effective tachyon potential
\begin{equation}
V (\phi)
= \sum_{n = 2}^{\infty} c_n (g \kappa)^{n-2}\phi^n
 = -\frac{1}{2}\phi^2 + g \kappa \phi^3+ \cdots
% \label{eq:}
\end{equation}
There are a number of ways in which we might try to calculate the
coefficients $c_n$ in this expansion.  Given a cubic potential
(\ref{eq:potential}) for $N$ fields, one approach to determining the
coefficients $c_n$ would be to write the quadratic equations of motion
for each of the $N -1$ fields which we wish to integrate out, expand
each field $\psi^i, i > 1$ as a formal power series in $\phi$, and
solve a linear system of equations at each order in $\phi$, plugging
the final results back into the cubic potential.  We will not directly
use this method, although it is technically equivalent to the
graphical approach we will use.

An alternative approach to calculating the coefficients in the
effective tachyon potential is to treat (\ref{eq:potential}) as the
action for a 0-dimensional field theory and to use 
Feynman diagrams to sum all terms contributing to a given coefficient
$c_n$.  This method was used by Kostelecky and Samuel in
\cite{ks-open} to calculate the level 2 and level 4 approximations to
the quartic coefficient $c_4$, which they had calculated exactly in
\cite{ks-exact}.  They found that at level 2, 72\% of the exact
coefficient is generated, and at level 4 this increases to 84\%.  This
calculation was extended in \cite{WT-SFT}, where contributions from
all fields up to level 20 were included, generating 96-97\% of the
exact term $c_4$.

We can use the exact cubic terms we have found up to level (10, 20) as
cubic interaction vertices in Feynman diagrams which we can then sum
to find the contributions to each of the coefficients $c_n$.  Because
the combinatorics of the Feynman diagrams grows exponentially, we will
find it useful to use an algebraic simplification to expedite the
summation over graphs.  The essence of the simplification we will use
is that because we are summing over planar tree graphs, we can write a
recursion relation relating the set of graphs $G_k$ with $k$ external
$\phi$ edges and single external $\psi^i$ edge to the sets of graphs
$G_{k'}$ with $k' < k$.  This relationship is indicated schematically
by
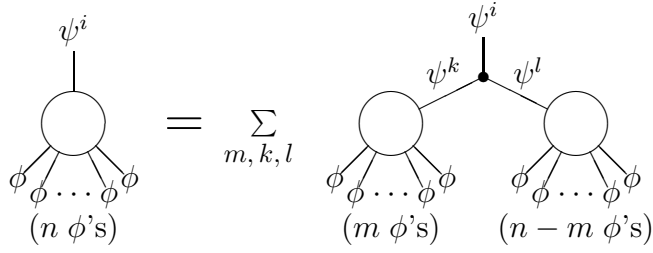
\[
\centering
\begin{picture}(200,100)(- 100,- 50)
\put(-150,-50){\begin{picture}(100, 100)(- 50,- 50)
\put(0,0){\circle{24}}
\put(0,12){\line(0,1){15}}
\put(8.48528,-8.48528){\line( 1, -1){10.607}}% length 15
\put( -8.48528,-8.48528){\line(-1, -1){10.607}}% length 15
\put(5.36656, -10.7331){\line( 1,-2){6.7082}}
\put( -5.36656, -10.7331){\line(-1,-2){6.7082}}
\put(-21.2132, -21.2132){\makebox(0,0){$\phi$}}
\put(21.2132, -21.2132){\makebox(0,0){$\phi$}}
\put(-13.4164,-26.8328){\makebox(0,0){$\phi$}}
\put(13.4164,-26.8328){\makebox(0,0){$\phi$}}
\put(0,-27){\makebox(0,0){$\cdots$}}
\put(0,-40){\makebox(0,0){($n\; \phi$'s)}}
\put(0,35){\makebox(0,0){$\psi^i$}}
\end{picture}}
\put(-50,0){\makebox(0,0){\LARGE $\; \;= \;\;\sum$}}
\put(-30,-12){\makebox(0,0){\footnotesize $m, k, l$}}
\put(30.7331, 5.36656){\line(2,1){24.2669}}
\put(55, 17.5){\line(2, -1){24.2669}}
\put(55, 17.5){\line(0, 1){15}}
\put(55, 17.5){\circle*{4}}
\put(55, 40){\makebox(0,0){$\psi^i$}}
\put(40,19){\makebox(0,0){$\psi^k$}}
\put(72,19){\makebox(0,0){$\psi^l$}}
\put(-30,-50){\begin{picture}(100, 100)(- 50,- 50)
\put(0,0){\circle{24}}
%\put(0,12){\line(0,1){15}}
\put(8.48528,-8.48528){\line( 1, -1){10.607}}% length 15
\put( -8.48528,-8.48528){\line(-1, -1){10.607}}% length 15
\put(5.36656, -10.7331){\line( 1,-2){6.7082}}
\put( -5.36656, -10.7331){\line(-1,-2){6.7082}}
\put(-21.2132, -21.2132){\makebox(0,0){$\phi$}}
\put(21.2132, -21.2132){\makebox(0,0){$\phi$}}
\put(-13.4164,-26.8328){\makebox(0,0){$\phi$}}
\put(13.4164,-26.8328){\makebox(0,0){$\phi$}}
\put(0,-27){\makebox(0,0){$\cdots$}}
\put(0,-40){\makebox(0,0){($m\; \phi$'s)}}
%\put(0,35){\makebox(0,0){$\psi^i$}}
\end{picture}}
\put(40,-50){\begin{picture}(100, 100)(- 50,- 50)
\put(0,0){\circle{24}}
%\put(0,12){\line(0,1){15}}
\put(8.48528,-8.48528){\line( 1, -1){10.607}}% length 15
\put( -8.48528,-8.48528){\line(-1, -1){10.607}}% length 15
\put(5.36656, -10.7331){\line( 1,-2){6.7082}}
\put( -5.36656, -10.7331){\line(-1,-2){6.7082}}
\put(-21.2132, -21.2132){\makebox(0,0){$\phi$}}
\put(21.2132, -21.2132){\makebox(0,0){$\phi$}}
\put(-13.4164,-26.8328){\makebox(0,0){$\phi$}}
\put(13.4164,-26.8328){\makebox(0,0){$\phi$}}
\put(0,-27){\makebox(0,0){$\cdots$}}
\put(0,-40){\makebox(0,0){($n -m\; \phi$'s)}}
%\put(0,35){\makebox(0,0){$\psi^i$}}
\end{picture}}
\end{picture}
\]
Algebraically, we can define an
$N$-dimensional vector $v_n^i$ for each $n$, representing the
summation over all graphs with $n$ external $\phi$ edges and a single
external $\psi^i$ (including the propagator for $\psi$).  We then have
\begin{eqnarray}
v_1^i & = &  \delta^i_1 \nonumber\\
v_n^i & = & \frac{3}{2} 
 \sum_{m = 1}^{n-1}  d^{ij}\, t_{jkl}\, \hat{v}_m^k \hat{v}_{n-m}^l
\label{eq:recursion}
\end{eqnarray}
where $d^{ij}$ is the inverse matrix to $d_{ij}$ and
\begin{equation}
\hat{v}_n^i = \left\{
\begin{array}{l}
0, \;\;\;\;\;i = 1 \; {\rm and} \; n > 1\\
v_n^i, \,\;\;\; {\rm otherwise}
\end{array}
\right.
\end{equation}
has been defined to project out internal $\phi$ edges.  In terms of
the vectors $v_n^i$, the coefficients $c_n$ are given by
\begin{equation}
c_n = \frac{1}{n}  v_{n-1}^1
% \label{eq:}
\end{equation}
In fact, the recursion relations (\ref{eq:recursion}) precisely encode
the relations we would find between the coefficients of the fields
$\psi^i$ expanded in powers of $\phi$ if we used the power series
approach to solving the $N -1$ quadratic equations term by term in
$\phi$ as described in the first paragraph of this subsection.  The
terms $v_n^i$ are precisely the coefficients of $\phi^n$ in the
expansion of the field $\psi^i$.

In any case, using this recursive formalism, the computation of $c_n$
becomes a polynomial-time algorithm taking time of order ${\cal O}
(N^3 n^2)$, instead of an exponentially hard algorithm such as we
would encounter if we tried to directly sum over all Feynman diagrams.
It may be helpful to illustrate this approach with a simple
example.  Consider  a single massive field $\psi$ which
couples to $\phi$ only through a term of the form $\phi^2 \psi$ in  the
potential
\begin{equation}
V = -\frac{1}{2}\phi^2 + g \kappa \phi^3
+\frac{1}{2} \psi^2 + g \kappa \left(
\frac{\gamma}{6}  \psi^3 + \beta \phi^2 \psi\right).
\label{eq:psi-potential}
\end{equation}
In this case we can explicitly solve for $\psi$
\begin{equation}
\psi = \frac{1}{g \kappa\gamma}
  \left( -1 + \sqrt{1-2 g^2\kappa^2\beta \gamma \phi^2} \right)
\label{eq:solve-psi}
\end{equation}
As mentioned above, we have chosen the branch of the square root which
gives $\psi = 0$ when $\phi = 0$.  Substituting (\ref{eq:solve-psi})
into (\ref{eq:psi-potential})
gives us the effective potential for $\phi$ on the branch containing
the unstable vacuum
\begin{equation}
V = -\frac{1}{2}\phi^2 + g \kappa \phi^3 +
\frac{1}{3 g^2 \kappa^2 \gamma^2} \left(1  -3g^2 \kappa^2 \beta \gamma \phi^2
-(1-2 g^2 \kappa^2\beta \gamma \phi^2)^{3/2} \right)
\label{eq:effective-psi}
\end{equation}
The coefficients in a power series expansion of this potential are
\begin{equation}
c_{2n} = -\frac{(2n-5)!!}{n!} \beta^n \gamma^{n-2}
\label{eq:coefficients-example}
\end{equation}
for $2n  \geq 4$.

To derive these coefficients using the recursive formalism
(\ref{eq:recursion}) we need the coefficients
\begin{eqnarray*}
d_{11} = -\frac{1}{2} &  &  t_{112} = \frac{\beta}{3} \\
d_{ 22} = \frac{1}{2} &  &  t_{222} = \frac{\gamma}{6} 
\end{eqnarray*}
which give the recursion relations
\begin{eqnarray*}
v^1_1 & = &  1\\
v^1_n & = &  -2 \beta \sum_{m = 1}^{n-1}  \hat{v}^1_{m}
\hat{v}^2_{n-m} = -2 \beta v^2_{n-1}, \;\;\;\;\; n > 1\\
v^2_1 & = &  0\\
v^2_2 & = &  \beta  \hat{v}^1_1 \hat{v}^1_1= \beta\\
v^2_n & = &  \frac{\gamma}{2}  \sum_{m = 1}^{ n-1}  v^2_mv^2_{n-m},
\;\;\;\;\; n > 2.
\end{eqnarray*}
We begin by solving for $v^2_n$.  These terms clearly vanish unless
$n$ is even.  Defining
\begin{equation}
w_m = v^2_{2m}
% \label{eq:}
\end{equation}
we have
\begin{eqnarray}
w_1 & = &  \beta \label{eq:recursion-w}\\
w_m & = &  \frac{\gamma}{ 2}  \sum_{k = 1}^{ m-1}  w_kw_{m-k}. \nonumber
\end{eqnarray}
Defining a generating function
\begin{equation}
f (x) = \sum_{m = 1}^{ \infty}  w_mx^{2m}
% \label{eq:}
\end{equation}
The equations (\ref{eq:recursion-w})
are equivalent to the quadratic equation
\begin{equation}
f = \frac{\gamma}{2}  f^2 + \beta x^2
% \label{eq:}
\end{equation}
with solution
\begin{equation}
f = \frac{1}{\gamma}  (1-\sqrt{1-2 \beta \gamma x^2})
= \sum_{m = 1}^{ \infty}  \frac{(2m-3)!!}{m!}  \beta^m
\gamma^{m-1} x^{2m}
% \label{eq:}
\end{equation}
So
\begin{equation}
c_{2n} = \frac{1}{2n}  (-2 \beta) w_{n-1}
=-\frac{(2n-5)!!}{n!} \beta^n \gamma^{n-2}
% \label{eq:}
\end{equation}
in agreement with (\ref{eq:coefficients-example}).

\subsection{Coefficients of effective tachyon potential}
\label{sec:coefficients}

We have used the method described in the previous subsection to
calculate the coefficients $c_{n}$ in the perturbative expansion of
the effective tachyon potential around the unstable vacuum for $n \leq
60$ in each of the level-truncated theories up to (10, 20).
In Table~\ref{t:coefficients} we have tabulated the successive
approximations to $c_n$ for a representative set of values of $n$ at
various levels.
\begin{table}[htp]
\begin{center}
\begin{tabular}{ || c || c |  c | c | c |c ||}
\hline
\hline
$n$ & (2, 6) & (4, 12) & (6, 18) & (8, 20) & (10, 20)\\
\hline
\hline
3 & $1
$ & $1
$ & $1
$ & $1
$ & $1$\\ \hline
4 & $-1.2592
$ & $-1.4724
$ & $-1.5562
$ & $-1.6004
$ & $-1.6276$\\ \hline
5 & $5.9478
$ & $7.8370
$ & $8.6398
$ & $9.0756
$ & $9.3479$\\ \hline
6 & $-27.909
$ & $-43.529
$ & $-50.768
$ & $-54.816
$ & $-57.353$\\ \hline
7 & $143.67
$ & $269.25
$ & $333.83
$ & $371.36
$ & $395.12$\\ \hline
8 & $-786.81
$ & $-1777.5
$ & $-2346.7
$ & $-2691.9
$ & $-2913.4$\\ \hline
9 & $4513.9
$ & $12323.
$ & $17345.
$ & $20529.
$ & $22606.$\\ \hline
10 & $-26845.
$ & $-88690.
$ & $-133179.
$ & $-162693.
$ & $-182299.$\\ \hline
20 & $-4.0710\cdot{10}^{12}
$ & $-9.2769\cdot{10}^{13}
$ & $-2.6957\cdot{10}^{14}
$ & $-4.5477\cdot{10}^{14}
$ & $-6.0780\cdot{10}^{14}$\\ \hline
30 & $-1.3415\cdot{10}^{21}
$ & $-2.1308\cdot{10}^{23}
$ & $-1.2050\cdot{10}^{24}
$ & $-2.8154\cdot{10}^{24}
$ & $-4.4920\cdot{10}^{24}$\\ \hline
40 & $-6.0173{10}^{29}
$ & $-6.6758\cdot{10}^{32}
$ & $-7.3563\cdot{10}^{33}
$ & $-2.3817\cdot{10}^{34}
$ & $-4.5372\cdot{10}^{34}$\\ \hline
50 & $-3.1889\cdot{10}^{38}
$ & $-2.4732\cdot{10}^{42}
$ & $-5.3126\cdot{10}^{43}
$ & $-2.3840\cdot{10}^{44}
$ & $-5.4231\cdot{10}^{44}$\\ \hline
\hline
\end{tabular}
\caption[x]{\footnotesize Level truncation approximations to
coefficients $c_n$ in effective tachyon potential}
\label{t:coefficients}
\end{center}
\end{table}
There are several observations we can make based on these results.  In
\cite{WT-SFT} successive approximations to the coefficient $c_4$ were
computed using fields at up to level 20.  It was found that while the
level truncation method gives a monotonic sequence of approximations
which seem to converge to the known exact value for this coefficient
$c_4 \approx -1.75$, the convergence to the asymptotic value is fairly
slow.  For $c_4$, the contribution from the set of graphs including at
least one field of level $k$ but no fields of level $k + 1$ decreases
as $k$ increases.  The same is true for successive approximations to
$c_n$ for small $n$, but as $n$ increases we find that at $n =8$, the
contribution from graphs with some fields of level 4 exceeds that of
graphs with only fields of level 2.  At $n =15$, the contribution from
graphs with some fields of level 6 exceeds that of graphs with fields
of level at most 4.  The corresponding thresholds where fields of
levels 8 and 10 dominate the contributions from lower levels are $n =
26$ and $n = 43$ respectively.  The fact that higher level fields
become more relevant for higher order terms in the effective potential
is a natural consequence of the tradeoff between the exponential
growth in the number of diagrams contributing to $c_n$ and the
suppression of higher order diagrams by $(g \kappa)^{n-2}$.  From this
behavior, however, we see that for large $n$ it will be necessary to
include fields of increasingly high level in order to have a good
approximation to the coefficients $c_n$.

\begin{figure}[htp]
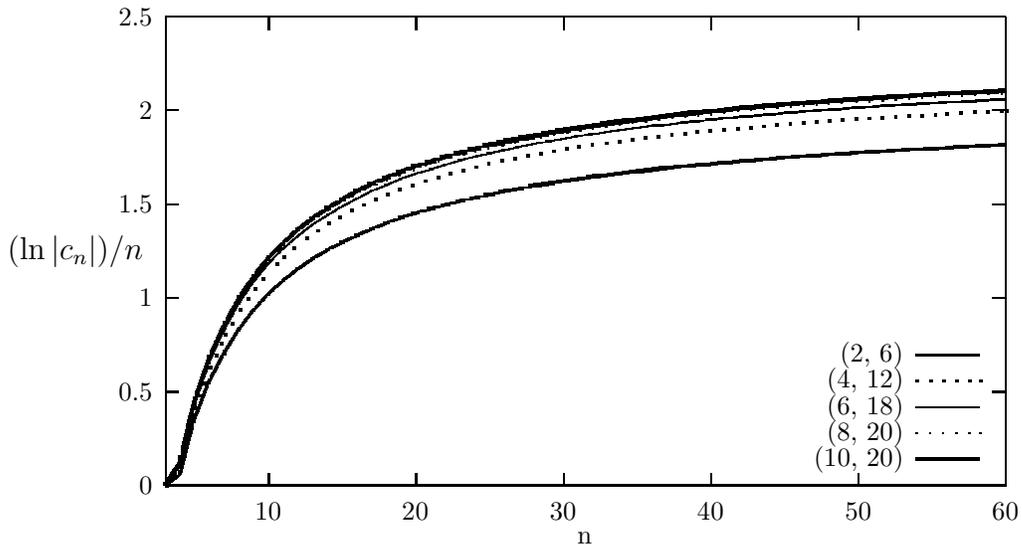

\begin{center}
% GNUPLOT: LaTeX picture
\setlength{\unitlength}{0.240900pt}
\ifx\plotpoint\undefined\newsavebox{\plotpoint}\fi
\sbox{\plotpoint}{\rule[-0.200pt]{0.400pt}{0.400pt}}%

\end{center}
\caption[x]{\footnotesize $(\ln | c_n |)/n$ for coefficients $c_n$ in
effective tachyon potential in different level truncations}
\label{f:coefficients}
\end{figure}

One interesting question which we can explore using our results for
the coefficients $c_n$ is the radius of convergence of the power
series expansion of the effective potential after truncating at a
fixed level.  Because of the square root branch cuts which arise from
the quadratic equations of motion for the fields, at some finite value
of $\phi$ the effective potential becomes singular for any given level
truncation.  For example, the effective potential arising from
(\ref{eq:psi-potential}) has a radius of convergence $r_c = 1/ (g
\kappa\sqrt{2 \beta \gamma})$, which is manifest in
(\ref{eq:effective-psi}) and which can be seen from the asymptotic
form of the coefficients $c_{2n} \sim \alpha n^\eta (\sqrt{2 \beta
\gamma})^{2n}$ where $\alpha, \eta$ are numerical constants.  It was found in
\cite{ks-open} that after truncation at level 2, there is a
singularity at $-\phi = r_c^{} \approx 0.35/g$.  This radius of
convergence can be seen from the coefficients of the effective
potential in the level 2 truncated theory, which for large $n$ go as
\begin{equation}
|c_n | \sim \alpha n^\eta \frac{1}{(g \kappa r_c)^n} 
\label{eq:c-asymptotic}
\end{equation}
A graph of $(\ln
|c_n |)/n$ as a function of $n$ 
is shown for each level truncation in
Figure~\ref{f:coefficients}.  
The asymptotic value of $(\ln |c_n|)/n$  gives the
(logarithm of the inverse of the) radius of convergence of the effective potential at
each level.  It is clear from the graph that this radius of
convergence is decreasing and seems to be approaching a nonzero
limiting value.  By matching the $c_n$ trajectories to a 3-parameter
family of functions of $n$ of the form (\ref{eq:c-asymptotic}) we can
find a close approximation to the radius of convergence at each of the
levels we have computed.  Table~\ref{t:radii} shows the approximate
radius of convergence at each level; we see that in the limiting
theory the radius of convergence approaches something like
\begin{equation}
r_c \approx 0.25/g\,.
\label{eq:radius}
\end{equation}
Because the signs of the coefficients $c_n$ are alternating, this
corresponds to a singularity in the effective potential at $g \phi
\sim -0.25$.  We study the behavior of the effective potential in the
vicinity of this singularity in more detail in section
\ref{sec:effective}.
\begin{table}[htp]
\begin{center}
\begin{tabular}{|| c  ||c | c |  c | c |  c  ||}
\hline
\hline
level & (2, 6)& (4, 12)& (6, 18) & (8, 20) & (10, 20)\\
\hline
$g r_c$ & $ \approx 0.345 $ & $ \approx 0.283 $ & $ \approx 0.265 $ & $ \approx
0.256 $ & $ \approx 0.252 $ \\ 
\hline
\hline
\end{tabular}
\caption[x]{\footnotesize Approximate radius of convergence of
effective potential in different level truncations}
\label{t:radii}
\end{center}
\end{table}

{}From this analysis of the perturbative effective potential using
string field theory we have found several things.  First, we find that
we can in principle determine any coefficient in the perturbative
expansion of the effective tachyon potential to an arbitrary degree of
accuracy with a finite calculation in string field theory, but that
the size of the calculation needed to determine the coefficients $c_n$
increases significantly as $n$ increases.  Second, we have found that
the resulting effective potential has a radius of convergence on the
order of (\ref{eq:radius}).  As we will discuss in the next section,
the stable vacuum of the theory lies well outside this radius, near
$\phi \approx 1.1/g$.  It is tempting to conclude from this that, even
in principle, the existence and structure of the stable vacuum of the
theory are fundamentally nonperturbative phenomena.

\section{The true vacuum and other nonperturbative features}
\label{sec:nonperturbative}

The appearance of a stable vacuum in the open bosonic string field
theory was first shown by Kostelecky and Samuel in \cite{ks-open}.
They found the vacuum in the $(2, 6)$ level-truncated theory.  In
later work, Kostelecky and Potting extended this analysis to the $(4,
12)$ level-truncated theory and reported that the energy density and
the values of the scalar fields in the new vacuum both seemed to be
rapidly converging as the level number was increased.  In
\cite{Sen-universality}, Sen argued that the energy difference between
the false and true vacua should precisely correspond to the tension of
the bosonic D25-brane.  Strong evidence for this conclusion was given
by Sen and Zwiebach in \cite{Sen-Zwiebach}, where they showed that at
level $(4, 8)$ the energy gap between the stable and unstable vacua
was $98.64\%$ of the D-brane energy.

In this section we use our results for the level-truncated string
field action to compute some nonperturbative features of the open
bosonic string.  In subsection (\ref{sec:vacuum}) we determine the
values of the scalar fields and the total energy of the system in the
stable vacuum  including fields of up to level 10.  At level (10, 20) we
find that the energy gap between the stable and unstable vacua
corresponds to $99.91\%$ of the D25-brane energy.  In subsection
\ref{sec:effective} we study the effective potential of the tachyon,
focusing on the branch structure of the effective
potential in the region $\phi < 0$.  

\subsection{The stable vacuum}
\label{sec:vacuum}

In order to find the stable vacuum of the level-truncated string field
theory action in the zero momentum sector it is necessary to solve a
system of $N$ coupled quadratic equations, where $N$ is the number of
scalar string fields involved.  In general, as $N$ becomes large it is
difficult to rapidly solve such a system of equations to a high degree
of precision.  Because we know which branch of each quadratic equation
the physical solution lies on, however, we can use an iterative
approximation algorithm to rapidly converge on the true vacuum.

As discussed in section \ref{sec:perturbative}, the branch of the
solution for each of the $N$
quadratic equations arising from the level-truncated string field
theory action is determined by the condition that the stable vacuum
lies on the same branch as the 
unstable vacuum
which has all fields vanishing.  Thus, the choice
of branch for each field is dictated by the sign of the quadratic term
in the string field action.  For most fields $\psi^i$,  therefore, the
value of the field can be expressed in terms of the remaining fields
$\psi^j, j \neq i$ through
\begin{equation}
\psi^i = \frac{-b + {\rm sign} (d_{ii}) \sqrt{b^2 -4ac}}{2a} 
\label{eq:solve1}
\end{equation}
where
\begin{eqnarray}
a & = &  3t_{iii}\nonumber\\
b & = &  \sum_{j \neq i} 6t_{iij} \psi^j + 2d_{ii}\label{eq:abc}\\
c & = &  \sum_{j, k \neq i} 3 t_{ijk} \psi^j \psi^k 
+\sum_{j \neq i}  2d_{ij} \psi^j\nonumber
\end{eqnarray}
There are some exceptions to this general rule, however.  For the
tachyon $\phi = \psi$ we choose the $+$ branch of the square root even
though the kinetic term is negative.  In addition, for most
fields with ghost excitations the quadratic term in the action is
off-diagonal.  For example, the level 4 fields $\psi^8, \psi^{10}$ are
coupled through the quadratic term $d_{8\, 10} =-3/2$.  For fields
such as this we use the equation of motion for field $\psi^i$ to solve
for the dual field $\tilde{\psi}^i$ which has ghost and antighost
modes exchanged  through $c_{-n} \leftrightarrow b_{-n}$.

We have used (\ref{eq:solve1}, \ref{eq:abc}) to solve iteratively for
the stable vacuum.  We begin by solving (\ref{eq:solve1}) for all
fields $\psi^i$, using zero for all fields appearing on the RHS.  We
then insert the first-round solutions for the fields back on the RHS
to solve again for all the $\psi^i$, and repeat this process many
times.  This algorithm is numerically very stable near the
nonperturbative vacuum and converges quite rapidly to a simultaneous
solution of all $N$ equations.  At level 6, for example, including all
interactions up to level 18, after 30 rounds of this procedure the
energy stabilizes to 10 digits, and after 50 rounds the values of all
fields stabilize to 10 digits.

In the units we are using here, the tension of the bosonic D25-brane
is \cite{Sen-universality}
\begin{equation}
T_{25} = \frac{1}{2 \pi^2 g^2} 
% \label{eq:}
\end{equation}
In Table~\ref{t:vacuum} we have tabulated the results of our
calculation of the exact vacuum in the theory truncated at various
levels.  
\begin{table}[htp]
\begin{center}
\begin{tabular}{|| c  || c |  c ||}
\hline
\hline
level & $ g \langle \phi \rangle$ & $V/T_{25}$\\
\hline
\hline
(0, 0) & 0.91236 & -0.68462\\
\hline
(2, 4) & 1.08318 &  -0.94855\\
(2, 6) & 1.08841 &  -0.95938\\
\hline
(4, 8) & 1.09633 &  -0.98640\\
(4, 12) & 1.09680 & -0.98782\\
\hline
(6, 12) & 1.09602 & -0.99514\\
(6, 18) & 1.09586 & -0.99518\\
\hline
(8, 16) & 1.09424 & -0.99777\\
(8, 20) & 1.09412 & -0.99793\\
\hline
(10, 20) & 1.09259 & -0.99912\\
\hline
\hline
\end{tabular}
\caption[x]{\footnotesize Tachyon VEV and vacuum energy in stable
vacua of level-truncated theory}
\label{t:vacuum}
\end{center}
\end{table}
The value of the tachyon is given in units of $1/g$ (with $\alpha' =
1$), and the vacuum energy difference from the unstable vacuum is
given as a proportion of the D25-brane tension.  The results at levels
(2, 4) and (2, 6) agree with those of \cite{ks-open}, and the results
at level (4, 8) agree with \cite{Sen-Zwiebach}.  

We see from Table~\ref{t:vacuum} that at level $(10, 20)$ the energy
gap between the unstable and stable vacua is $99.91\%$ of the
D25-brane tension.  This seems to affirm the prediction of Sen in
\cite{Sen-universality} beyond any reasonable doubt.  It is
interesting to note that the error in the vacuum energy $(1+V/T_{25})$
is multiplied by approximately $1/3 \approx \kappa$ as each new level
is added.  It would be nice to have a theoretical explanation for this
rate of convergence.

In \cite{ks-open} it was found that the vacuum energy and tachyon
expectation values change very little between the (2, 4) and (2, 6)
level truncations.  In \cite{Sen-Zwiebach} a calculation was performed
at level (4, 8) giving $98.6\%$ of the D-brane tension.  The
improvement on this result given by including level 10 and 12
interactions between level 4 fields is fairly small.  We have found
that this pattern persists at higher level.  The results at level (6,
18) are not very different from those found at level (6, 12), and the
results at level (8, 20) are very close to those at level (8, 16).
If, however, we drop cubic interactions at the same level as the
quadratic interactions at that level, for example by truncating all
cubic interactions to level (6, 6), the energy of the stable vacuum
varies much more wildly and can even decrease below $-T_{25}$.  For
this reason, we do not think that it would be useful to consider
including higher level fields unless the level of cubic interactions
calculated could be increased to at least twice the level number.
Using our rather inefficient program it would take a significant
amount of computer time to go to level (12, 24) on a standard desktop
machine.  We discuss in the last section the prospects for continuing
these numerical studies to higher levels of truncation of the full
theory.

As we can see from Table~\ref{t:vacuum}, the expectation value of the
tachyon in the stable vacuum converges fairly quickly as the level number
is increased.  The same is true of all the scalar fields; in Appendix
A the values of the scalar fields at levels $\leq 6$ are given for the
level $(4, 12)$, $(6, 18)$ and $(10, 20)$ truncations.  The  tachyon
field itself converges to a value near
\begin{equation}
g
\langle \phi \rangle \approx
1.09
% \label{eq:}
\end{equation}

\begin{figure}
\begin{center}
% GNUPLOT: LaTeX picture
\setlength{\unitlength}{0.240900pt}
\ifx\plotpoint\undefined\newsavebox{\plotpoint}\fi
\sbox{\plotpoint}{\rule[-0.200pt]{0.400pt}{0.400pt}}%
\begin{picture}(1500,900)(0,0)
\font\gnuplot=cmr10 at 10pt
\gnuplot
\sbox{\plotpoint}{\rule[-0.200pt]{0.400pt}{0.400pt}}%
\put(181.0,123.0){\rule[-0.200pt]{4.818pt}{0.400pt}}
\put(161,123){\makebox(0,0)[r]{0.18}}
\put(1440.0,123.0){\rule[-0.200pt]{4.818pt}{0.400pt}}
\put(181.0,215.0){\rule[-0.200pt]{4.818pt}{0.400pt}}
\put(161,215){\makebox(0,0)[r]{0.185}}
\put(1440.0,215.0){\rule[-0.200pt]{4.818pt}{0.400pt}}
\put(181.0,307.0){\rule[-0.200pt]{4.818pt}{0.400pt}}
\put(161,307){\makebox(0,0)[r]{0.19}}
\put(1440.0,307.0){\rule[-0.200pt]{4.818pt}{0.400pt}}
\put(181.0,399.0){\rule[-0.200pt]{4.818pt}{0.400pt}}
\put(161,399){\makebox(0,0)[r]{0.195}}
\put(1440.0,399.0){\rule[-0.200pt]{4.818pt}{0.400pt}}
\put(181.0,492.0){\rule[-0.200pt]{4.818pt}{0.400pt}}
\put(161,492){\makebox(0,0)[r]{0.2}}
\put(1440.0,492.0){\rule[-0.200pt]{4.818pt}{0.400pt}}
\put(181.0,584.0){\rule[-0.200pt]{4.818pt}{0.400pt}}
\put(161,584){\makebox(0,0)[r]{0.205}}
\put(1440.0,584.0){\rule[-0.200pt]{4.818pt}{0.400pt}}
\put(181.0,676.0){\rule[-0.200pt]{4.818pt}{0.400pt}}
\put(161,676){\makebox(0,0)[r]{0.21}}
\put(1440.0,676.0){\rule[-0.200pt]{4.818pt}{0.400pt}}
\put(181.0,768.0){\rule[-0.200pt]{4.818pt}{0.400pt}}
\put(161,768){\makebox(0,0)[r]{0.215}}
\put(1440.0,768.0){\rule[-0.200pt]{4.818pt}{0.400pt}}
\put(181.0,860.0){\rule[-0.200pt]{4.818pt}{0.400pt}}
\put(161,860){\makebox(0,0)[r]{0.22}}
\put(1440.0,860.0){\rule[-0.200pt]{4.818pt}{0.400pt}}
\put(181.0,123.0){\rule[-0.200pt]{0.400pt}{4.818pt}}
\put(181,82){\makebox(0,0){2}}
\put(181.0,840.0){\rule[-0.200pt]{0.400pt}{4.818pt}}
\put(501.0,123.0){\rule[-0.200pt]{0.400pt}{4.818pt}}
\put(501,82){\makebox(0,0){4}}
\put(501.0,840.0){\rule[-0.200pt]{0.400pt}{4.818pt}}
\put(821.0,123.0){\rule[-0.200pt]{0.400pt}{4.818pt}}
\put(821,82){\makebox(0,0){6}}
\put(821.0,840.0){\rule[-0.200pt]{0.400pt}{4.818pt}}
\put(1140.0,123.0){\rule[-0.200pt]{0.400pt}{4.818pt}}
\put(1140,82){\makebox(0,0){8}}
\put(1140.0,840.0){\rule[-0.200pt]{0.400pt}{4.818pt}}
\put(1460.0,123.0){\rule[-0.200pt]{0.400pt}{4.818pt}}
\put(1460,82){\makebox(0,0){10}}
\put(1460.0,840.0){\rule[-0.200pt]{0.400pt}{4.818pt}}
\put(181.0,123.0){\rule[-0.200pt]{308.111pt}{0.400pt}}
\put(1460.0,123.0){\rule[-0.200pt]{0.400pt}{177.543pt}}
\put(181.0,860.0){\rule[-0.200pt]{308.111pt}{0.400pt}}
%\put(-5,491){\makebox(0,0){$(\ln |c_n|)/n$}}
\put(820,41){\makebox(0,0){Truncation level}}
\put(181.0,123.0){\rule[-0.200pt]{0.400pt}{177.543pt}}
\put(1300,205){\makebox(0,0)[r]{u}}
\put(1320.0,205.0){\rule[-0.200pt]{24.090pt}{0.400pt}}
\put(181,311){\usebox{\plotpoint}}
\multiput(181.00,311.58)(0.561,0.500){567}{\rule{0.549pt}{0.120pt}}
\multiput(181.00,310.17)(318.860,285.000){2}{\rule{0.275pt}{0.400pt}}
\multiput(501.00,596.58)(1.418,0.499){223}{\rule{1.233pt}{0.120pt}}
\multiput(501.00,595.17)(317.441,113.000){2}{\rule{0.616pt}{0.400pt}}
\multiput(821.00,709.58)(2.716,0.499){115}{\rule{2.263pt}{0.120pt}}
\multiput(821.00,708.17)(314.304,59.000){2}{\rule{1.131pt}{0.400pt}}
\multiput(1140.00,768.58)(4.749,0.498){65}{\rule{3.865pt}{0.120pt}}
\multiput(1140.00,767.17)(311.979,34.000){2}{\rule{1.932pt}{0.400pt}}
\put(181,311){\raisebox{-.8pt}{\makebox(0,0){$\circ$}}}
\put(501,596){\raisebox{-.8pt}{\makebox(0,0){$\circ$}}}
\put(821,709){\raisebox{-.8pt}{\makebox(0,0){$\circ$}}}
\put(1140,768){\raisebox{-.8pt}{\makebox(0,0){$\circ$}}}
\put(1460,802){\raisebox{-.8pt}{\makebox(0,0){$\circ$}}}
\put(1370,205){\raisebox{-.8pt}{\makebox(0,0){$\circ$}}}
\sbox{\plotpoint}{\rule[-0.500pt]{1.000pt}{1.000pt}}%
\put(1300,164){\makebox(0,0)[r]{v}}
\multiput(1320,164)(20.756,0.000){5}{\usebox{\plotpoint}}
\put(1420,164){\usebox{\plotpoint}}
\put(181,524){\usebox{\plotpoint}}
\multiput(181,524)(20.352,4.070){16}{\usebox{\plotpoint}}
\multiput(501,588)(20.733,0.972){16}{\usebox{\plotpoint}}
\multiput(821,603)(20.754,0.260){15}{\usebox{\plotpoint}}
\multiput(1140,607)(20.755,0.130){15}{\usebox{\plotpoint}}
\put(1460,609){\usebox{\plotpoint}}
\put(181,524){\makebox(0,0){$\bullet$}}
\put(501,588){\makebox(0,0){$\bullet$}}
\put(821,603){\makebox(0,0){$\bullet$}}
\put(1140,607){\makebox(0,0){$\bullet$}}
\put(1460,609){\makebox(0,0){$\bullet$}}
\put(1370,164){\makebox(0,0){$\bullet$}}
\end{picture}

\end{center}
\caption[x]{\footnotesize Expectation values of fields $u = -\psi^3$
and $v = \psi^2/\sqrt{52}$ for $g = 2$ in different level truncations}
\label{f:uv}
\end{figure}
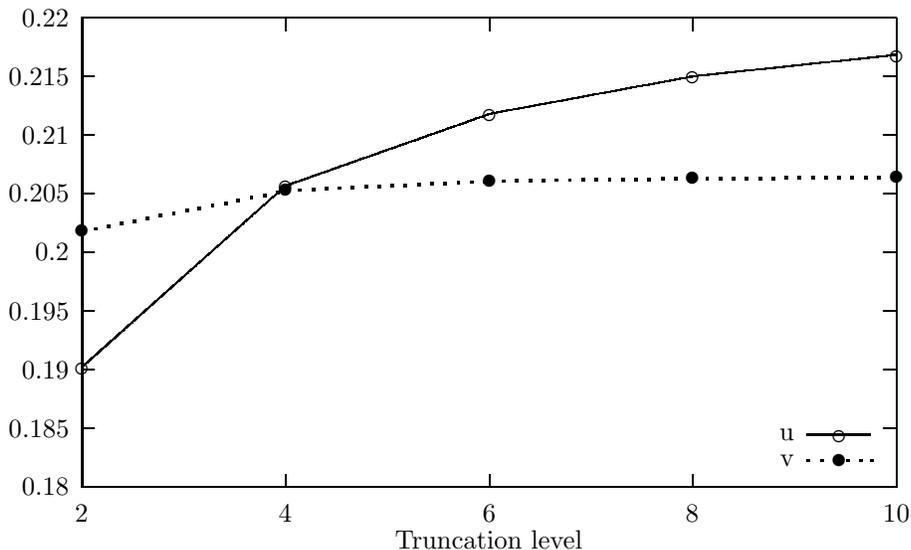

In \cite{Sen-Zwiebach} Sen and Zwiebach observed that the level 2 fields with
canonical normalization
\begin{eqnarray*}
u & = &  -\psi^3\\
v & = &  \frac{1}{2 \sqrt{13}}  \psi^2
\end{eqnarray*}
take almost identical expectation values in the level $(4, 8)$
truncation.  This led these authors to conjecture that in the exact
theory these fields would take identical expectation values, possibly
due to some hidden symmetry in the theory.  Checking this relationship
in the vacua of the higher level truncations, we find that this
conjecture is not supported.  While at level $(4, 8)$ these fields
differ by about $0.1\%$, at level $(10, 20)$ the difference has grown
to over $5\%$.  The values of these fields in the different
level-truncated vacua are graphed in Figure~\ref{f:uv}, using the
value  $g = 2$ to conform with the conventions of \cite{Sen-Zwiebach}.
The failure of this equality to hold in the asymptotic theory
indicates that it may be more difficult than previously thought to
implement the suggestion made in \cite{Sen-Zwiebach} of finding an
exact solution for the nonperturbative vacuum in the full theory using
a hidden symmetry relating fields of this type.

As a check on our results, we can verify that the vacuum expectation
values of the fields $\psi^i$ are such that the nonperturbative vacuum
lies in the truncated Hilbert space $H_1$ described in
\cite{Sen-universality}.  For example, the level 4 fields should obey
the linear relation
\begin{equation}
\langle \psi^4 \rangle -
2\langle \psi^5 \rangle -
4\langle \psi^6 \rangle = 0
% \label{eq:}
\end{equation}
in the vacuum.  We have checked this relation and other analogous
relations for higher level fields and find that they are satisfied up
to the level of accuracy (10 significant digits) to which we have
calculated the vacuum expectation values.

Now that we have identified the stable vacuum of the theory after
truncating at levels up to (10, 20), a natural next step is to
investigate the structure of the theory around this stable vacuum.
The stable vacuum should correspond to the empty vacuum of the closed
bosonic string theory, with modes corresponding to the open string
fields on the D25-brane decoupling from the theory.  Of particular
interest in this regard is the fate of the U(1) vector field $A^\mu$
on the brane \cite{Srednicki-IIB,Witten-K,Yi-membranes}.  Some
progress in understanding the structure of the theory around the
stable vacuum was made in \cite{ks-open}.  It would be interesting to
study this question further and to investigate the spectrum of modes
around the stable vacuum using the more detailed picture we have given
of the expectation values of the scalar string fields in this vacuum.
We leave these questions to further work.

\subsection{Effective tachyon potential}
\label{sec:effective}

In section \ref{sec:coefficients} we studied the effective potential
of the tachyon through its power series expansion around the unstable
vacuum $\psi^i = 0$.  We found that the perturbation series had a finite
radius of convergence in each of the level-truncated theories,
indicating a possible singularity near $\phi \approx -0.25/g$.  In
this section we study the nonperturbative effective potential, and
investigate the branch structure of the potential near the singularity.

For a fixed value of the tachyon field, there are in general many
solutions of the equations for the remaining fields $\psi^i, i > 1$
which correspond to different branches of the effective potential.
The perturbative expansion describes only one branch of the effective
potential, the one which is connected to the unstable vacuum.  We will
refer to this as branch 1.  For levels higher than (2,4), we cannot
exactly integrate out all the non-tachyonic fields, so we are forced
to use numerical methods to study the effective potential outside its
radius of convergence or on other branches.  We have used Newton's
method to find the zeros of the partial derivatives of the
potential. The choice of initial values for the fields determines on
which branch the algorithm will converge.  At mass level 2, we can
find all the solutions, and thus determine all the branches. We can
then use the field values from these branches as initial values for
the algorithm at higher levels in order to stay on the same branch.

\begin{figure}
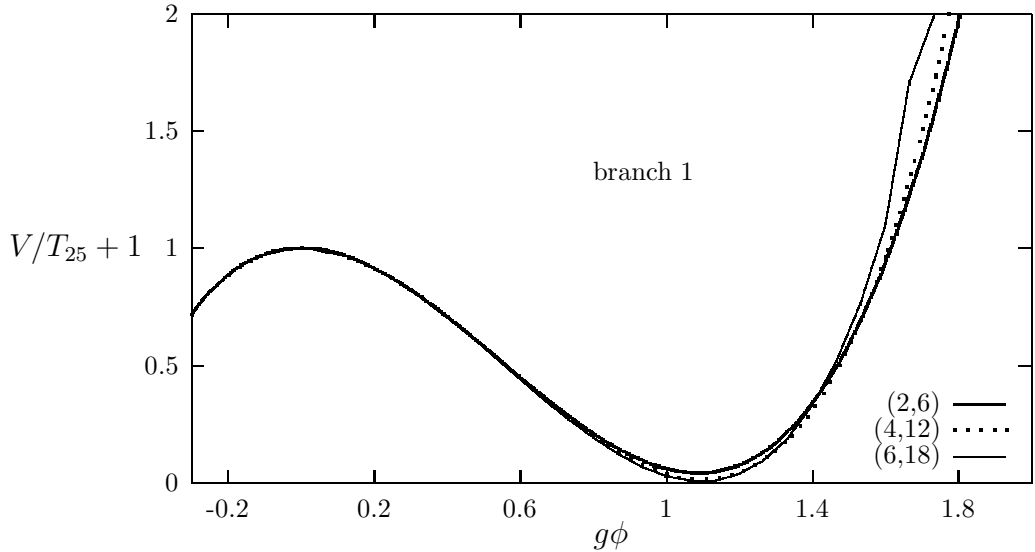

\begin{center}
% GNUPLOT: LaTeX picture
\setlength{\unitlength}{0.240900pt}
\ifx\plotpoint\undefined\newsavebox{\plotpoint}\fi

\end{center}
\caption[x]{Branch 1 of the effective tachyon potential at different
truncation 
levels}
\label{branch1}
\end{figure}

In Figure~\ref{branch1}, we have plotted branch 1 of the effective
potential at levels (2,6), (4,12) and (6,18). At level (4,12), our
algorithm stops converging after $g\phi \approx 1.8$.  At level
(6,18), the algorithm becomes unstable after $g \phi \approx
1.6$. This may indicate either that the branch ends at that point or
that it meets one or more other branches which play the role of
attractors, making it difficult to converge to the chosen branch.  A
similar breakdown of convergence also happens at level (2, 6) around
$g \phi \approx 6$, where a new pair of branches appear
\cite{ks-open}.

A particularly interesting physical question is to determine the
structure of the tachyon effective potential for negative values of
$\phi$.  In the level 0 truncated theory, this potential decreases to
$-\infty$ as $\phi \rightarrow -\infty$.  If this behavior is not
modified by higher level corrections, it poses the natural question of
why the D25-brane would choose to condense to the stable vacuum rather
than the runaway solution at negative $\phi$.

\begin{figure}
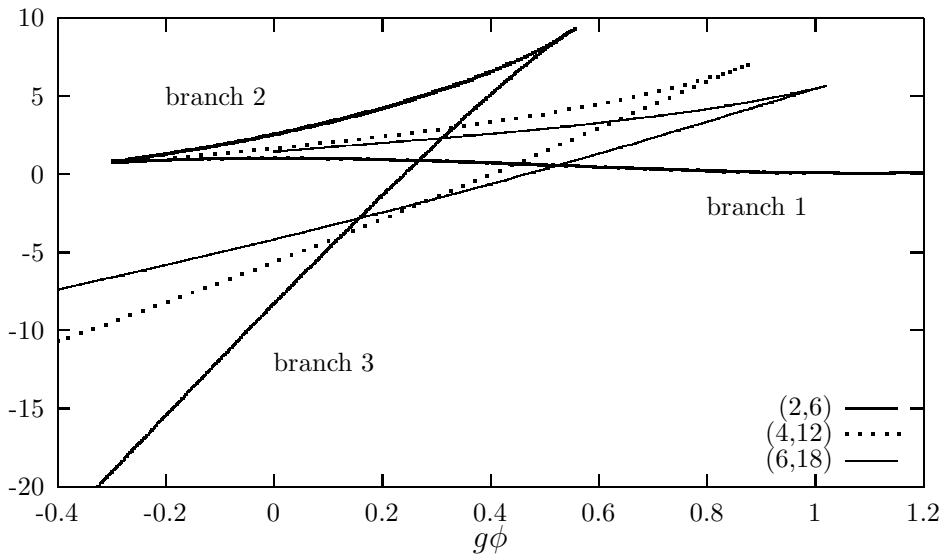

\begin{center}
% GNUPLOT: LaTeX picture
\setlength{\unitlength}{0.240900pt}
\ifx\plotpoint\undefined\newsavebox{\plotpoint}\fi
\sbox{\plotpoint}{\rule[-0.200pt]{0.400pt}{0.400pt}}%

\end{center}
\caption[x]{Structure of branches 1, 2, and 3 at different levels}
\label{branches123}
\end{figure}

When we attempt to continue branch 1 of the effective potential to
include arbitrary negative values of the tachyon field, we find that
at each level our algorithm stops converging very near the radius of
convergence given in Table \ref{t:radii}, which we calculated using
the perturbative expansion coefficients of the effective potential.
In \cite{ks-open}, Kostelecky and Samuel studied the effective
potential at mass level 2 near this point. They found that the branch
connected to the unstable and stable vacua (branch 1) ends at a
singularity at $g \phi \approx -0.35$ and, at that point, meets
another branch (branch 2) which continues upward for increasing values
of $\phi$. This branch then meets a third branch (branch 3) which
continues downward for decreasing values of $\phi$.  We have studied
the branch structure of the effective potential in level truncations
(4, 12) and (6, 18) and we find precisely the same branch structure,
although the shape of the branches changes noticeably at each level.
In Figure~\ref{branches123}, a graph is given of the branch structure
of the effective potential at each of these levels.  We observe that
branch 3 becomes less steep as we go at higher levels, and that it
begins to approach the singular point where branch 1 meets branch 2.
While branch 3 does not seem to approach the singular point of branch
1 exponentially quickly, its rate of approach seems compatible with
the rate of convergence of the coefficients in the perturbative
expansion of the tachyon effective potential.  Although we only have
limited evidence for this suggestion at this time, we conjecture that
in the limit of the full string field theory, branch 3 precisely
intersects the singular point connecting branches 1 and 2.  The
physical effect of this would be to create a critical point near $g
\phi \approx -0.25$ at which there would be a higher-order phase
transition in the tachyon effective potential.  Unlike the effective
potential on branch 1, the structure of the effective potential on
branch 3 changes dramatically with each additional level of string
fields which are included.  Thus, we cannot trust our low-level
truncation of the theory to accurately describe the effective
potential beyond the critical point.  It is natural to speculate that
the effective potential becomes nonnegative (relative to the energy of
the stable vacuum) for all values of $\phi$ in the full string field
theory.

It would be very nice to have some better understanding of the physics
of open bosonic string theory beyond the critical point; there must
also be a natural physical explanation for the existence of the
critical point, which might be understood from the point of view of
bosonic D-branes.  We leave these questions for further research, but
we conclude this section by pointing out one possible amusing
scenario: it is quite possible that the numerical difficulties we have
encountered in extending branch 1 of the effective potential near
$g\phi \approx 1.6$ indicate a second critical point at approximately
the same energy as that we have found at $g \phi \approx -0.25$.  If
this is indeed the case, it may be that these two critical points
should be physically identified in the full theory, so that the
tachyon effective potential will essentially become (in some
coordinates) periodic, and perhaps even smooth.  This would provide a
satisfying resolution to the question of what happens when the tachyon
rolls in the negative direction: it would simply bring the system to
the same stable vacuum as if it rolled in the positive direction.
More evidence is needed before this possibility can be taken
seriously, but it would provide a nice scenario in which the effective
potential for the tachyon has a unique global minimum corresponding to
the stable vacuum, just as occurs for the superstring \cite{bsz}.
One potentially serious drawback to this scenario, however, is that it
seems to predict the existence of a stable 24-brane kink for whose
existence there is no evidence in the bosonic string theory.

\section{Conclusions}

In this paper we have used the level-truncation approach to string
field theory to perform a detailed study of the tachyonic instability
of the open bosonic string.  We have calculated the cubic string field
theory potential up to terms of mass level 20, including fields up to
level 10.  We have used these results to analyze the effective tachyon
potential and the nonperturbative stable vacuum of the system.  Our
results for the perturbative form of the effective potential indicate
that the radius of convergence of this potential decreases to an
apparently finite value as higher level fields are included,
indicating a critical point in the effective potential near $\phi
\approx -0.25/g$.  The stable vacuum lies well outside this radius of
convergence, although the critical point and the stable vacuum lie on
opposite sides of the unstable vacuum so there is no phase transition
between the two vacua.  We have found that while the coefficients in
the effective potential converge relatively slowly using the level
truncation method, the energy of the stable vacuum solution and the
values of the fields in this vacuum converge much more quickly.  We
found that the discrepancy between the exact D-brane tension and the
vacuum energy calculated in the truncated string field theory
decreases by approximately a factor of $1/3$ when string fields at
each additional mass level are included, although we do not have a
theoretical explanation for this rate of convergence.

It is perhaps somewhat surprising that this method converges more
rapidly for the vacuum energy calculation, which is a truly
nonperturbative feature of the system, than for the coefficients of
the effective potential, which are in principle computable using
perturbative methods.  This finding seems to indicate that string
field theory has the potential to be an extremely useful tool in
studying detailed aspects of nonperturbative string physics.  Even if
it is not possible to find an exact analytic solution of string field
theory describing the nonperturbative vacuum, it would be of great
interest to carry out a more detailed analysis of the asymptotic
properties of the level truncation approach.

The methods used in this paper give us a fairly complete picture of
the behavior of the effective tachyon potential when $ -0.25 < g \phi
< 1.5$.  We have found that a critical point appears in the tachyon
potential near $g \phi \approx -0.25$, beyond which it is difficult to
precisely determine the physics.  We have speculated that the
effective potential stays above the stable vacuum energy for negative
$\phi$ beyond the critical point, but we do not yet have conclusive
evidence for this conclusion.  It would be very nice to have a better
understanding of the behavior of the theory in the regime where the
tachyon is large and negative.  We have outlined one possible
scenario, in which the tachyon effective potential becomes periodic
and the stable vacuum appears at the unique minimum of this potential.
If this scenario is realized it would provide a simple answer to the
question of how the theory behaves if the tachyon chooses to roll in
the negative direction.  This would also provide a picture in which
unstable bosonic D-branes could naturally be interpreted as
sphalerons, as advocated in the supersymmetric case in \cite{hhk}.

There are many directions in which it would be interesting to extend
the work in this paper.  One obvious question is to ask how far it is
possible to extend the level truncation method in the open bosonic
string field theory.  For the results in this paper where we included
fields at level 10 we have computed 138,202 distinct cubic vertices.
Because many of these vertices involve hundreds of possible index
contraction combinations, this computation is already rather
time-consuming using a higher-level symbolic manipulation program such
as {\it Mathematica} on a desktop PC.  To continue to level (12, 24),
over a million vertices (1,381,097) would be needed, each involving
hundreds or thousands of index contractions.  With a more efficient
program and a powerful computer, it might be possible to push as far
as level (16, 32) or higher in the forseeable future if there are
physical questions of sufficient interest to motivate further
development in this direction (at this level there are over $10^8$
cubic vertices).  Although the number of vertices grows as the cube of
the number of fields at the
level at which the theory is truncated, both the number of fields
and the complexity of computing each vertex grows exponentially, so
that unless a better theoretical understanding of the structure of the
theory is developed it will probably never be feasible to perform
calculations in this theory beyond interactions of total level 40 or
so.

There are a number of other questions of significant interest which
could be investigated using truncations of bosonic string field theory
at levels considered in this paper.  As discussed in section
\ref{sec:vacuum} it is clearly important to attain a better
understanding of the structure of the theory around the
nonperturbative vacuum.  While some progress was made in this
direction in \cite{ks-open}, we do not yet have a very clear picture
of how the open string fields such as the U(1) gauge field decouple
from the theory in the stable vacuum.  Another question of interest
which can be addressed in the bosonic theory is the existence of vacua
which break Lorentz symmetry.  Such vacua have been shown to exist and
to converge up to level (6, 18) truncations of the theory
\cite{ks-open,Kostelecky-Potting}.  It would be nice to have a better
theoretical understanding of the significance of these vacua.

A related question which can be addressed using open bosonic string
field theory is that of describing the structure of D$p$-branes with
$p < 25$ as unstable configurations of open string fields.  Solutions
of this type have been found in conformal field theory
\cite{cklm,Polchinski-Thorlacius,fsw,Recknagel-Schomerus}.  The
possibility of realizing a bosonic D$(p-1)$-brane as a tachyonic lump
on a D$p$-brane was discussed in \cite{Sen-descent}.  The use of the
level truncation approach to string field theory in this context was
suggested in \cite{Sen-Zwiebach}, and was implemented by Harvey and
Kraus \cite{Harvey-Kraus}, who found numerical evidence for
D$p$-branes with $p > 18$ as lumps on the D25-brane after including
effects of some level 2 fields.  It would be very interesting to
extend this work to higher level truncations and to see whether the
physics of all D$p$-branes is indeed accurately described in
level-truncated string field theory.

The open bosonic string is an interesting model in which string field
theory is particularly simple.  Studying this theory further may allow
us to refine our understanding of certain qualitative aspects of
D-brane physics and tachyonic instabilities.  Nonetheless, if string
theory is ever to make concrete contact with observable phenomenology
it will almost certainly be in the context of a supersymmetric theory.
Given that the open string field theory seems to be able to access
interesting nonperturbative structure in string theory, it is clearly
of substantial interest to develop an equally systematic approach to
performing nonperturbative calculations using superstring field
theory.  While Witten's formulation of open string field theory can be
extended to the supersymmetric theory
\cite{Witten-SFT-2,Gross-Jevicki-3}, this formalism may be problematic
due to the necessity of considering higher order contact interactions
\cite{Wendt,Greensite-Klinkhamer}.  Several alternative formulations
of open superstring field theory have been suggested
\cite{amz,pty,Berkovits-general}.  Recently Berkovits has used his
alternative formulation of the supersymmetric open string field theory
\cite{Berkovits-general} to carry out a first approximation of the
energy gap in a tachyonic brane system, and has found that 60\% of the
brane energy is reproduced even in the first truncation of the theory
\cite{Berkovits-tachyon}.  This formalism has been more explicitly
developed and the calculation of the energy gap has been extended by
Berkovits, Sen and Zwiebach \cite{bsz}, who showed that including the
three scalar fields at the next relevant level gives 85\% of the brane
energy.  If explicit calculations in open superstring field theory can
indeed be systematically carried out in high-level truncations as we
have done for the bosonic theory, this promises to provide an exciting
new tool for investigating in detail nonperturbative issues in string
theory such as the description of non-BPS D-branes as brane-antibrane
bound states \cite{Sen-stable,Sen-tachyon,Bergman-Gaberdiel} and the
associated connection between D-brane charges and K-theory
\cite{Witten-K,Horava-K}.

\section*{Acknowledgements}

We would particularly like to thank Barton Zwiebach for helpful
discussions and explanations in the course of this work.  Thanks also
to Amihay Hanany, Yang-Hui He, Andreas Karch, Joe Minahan and Leonardo
Rastelli for helpful conversations.  The work of NM was supported in
part by the DOE through contract \#DE-FC02-94ER40818.  The work of WT
was supported in part by the A.\ P.\ Sloan Foundation and in part by
the DOE through contract \#DE-FC02-94ER40818.

%%%%%%%%%%%%%%%%%%%%%%
\newpage
\appendix

\section{Table of scalar states at levels $\leq 6$}

The following table describes the scalar states at levels $0, 2, 4$
and 6.
The SZ column relates these fields to the background-independent
fields used in \cite{Sen-Zwiebach}.  $m$ and $g$ are the indices of
the matter and ghost states into which each scalar decomposes through
(\ref{eq:tensor-decomposition}).   $g \langle \psi^i
\rangle_{L}$ denote  expectation values of the scalar
fields in  level truncations (4, 12), (6, 18) and (10, 20).

\begin{center}
\begin{tabular}{| r | r | r | r | r | r | r | r |}
\hline
\hline
$\psi^i$ & SZ & state &  $m$ &  $g$ & $g\langle \psi^i \rangle_4 $
&$g\langle \psi^i \rangle_6$
&$ g\langle \psi^i \rangle_{10}$\\
\hline
\hline
$\psi^{1}$ & $\phi$ & 
$
\,
\,| 0 \rangle $
&1&1&
 1.09680 & 
 1.09586 & 
 1.09259 
\\
\hline
$\psi^{2}$ & $v/\sqrt{52}$ & 
$
(\alpha_{-1}\cdot \alpha_{-1})
\,
\,| 0 \rangle $
&2&1&
 0.05692 & 
 0.05714 & 
 0.05723 
\\
$\psi^{3}$ & $-u$ & 
$
\,
b_{-1}
c_{-1}
\,| 0 \rangle $
&1&2&
-0.41135 & 
-0.42363 & 
-0.43372 
\\
\hline
$\psi^{4}$ & $A + B$ & 
$
(\alpha_{-1}\cdot \alpha_{-3})
\,
\,| 0 \rangle $
&4&1&
-0.01142 & 
-0.01146 & 
-0.01148 
\\
$\psi^{5}$ & $A/2$ & 
$
(\alpha_{-2}\cdot \alpha_{-2})
\,
\,| 0 \rangle $
&5&1&
-0.00512 & 
-0.00511 & 
-0.00509 
\\
$\psi^{6}$ & $B/4$ & 
$
(\alpha_{-1}\cdot \alpha_{-1})
(\alpha_{-1}\cdot \alpha_{-1})
\,
\,| 0 \rangle $
&6&1&
-0.00029 & 
-0.00031 & 
-0.00032 
\\
$\psi^{7}$ & $-F$ & 
$
(\alpha_{-1}\cdot \alpha_{-1})
\,
b_{-1}
c_{-1}
\,| 0 \rangle $
&2&2&
 0.00686 & 
 0.00740 & 
 0.00786 
\\
$\psi^{8}$ & $-C$ & 
$
\,
b_{-1}
c_{-3}
\,| 0 \rangle $
&1&5&
 0.11242 & 
 0.11478 & 
 0.11654 
\\
$\psi^{9}$ & $E$ & 
$
\,
b_{-2}
c_{-2}
\,| 0 \rangle $
&1&6&
 0.06621 & 
 0.06813 & 
 0.06990 
\\
$\psi^{10}$ & $D$ & 
$
\,
b_{-3}
c_{-1}
\,| 0 \rangle $
&1&7&
 0.03747 & 
 0.03826 & 
 0.03885 
\\
\hline
$\psi^{11}$ &  & 
$
(\alpha_{-1}\cdot \alpha_{-5})
\,
\,| 0 \rangle $
&10&1&
  & 
 0.00349 & 
 0.00350 
\\
$\psi^{12}$ &  & 
$
(\alpha_{-2}\cdot \alpha_{-4})
\,
\,| 0 \rangle $
&11&1&
  & 
 0.00293 & 
 0.00291 
\\
$\psi^{13}$ &  & 
$
(\alpha_{-3}\cdot \alpha_{-3})
\,
\,| 0 \rangle $
&12&1&
  & 
 0.00144 & 
 0.00143 
\\
$\psi^{14}$ &  & 
$
(\alpha_{-1}\cdot \alpha_{-3})
(\alpha_{-1}\cdot \alpha_{-1})
\,
\,| 0 \rangle $
&13&1&
  & 
 0.00028 & 
 0.00028 
\\
$\psi^{15}$ &  & 
$
(\alpha_{-2}\cdot \alpha_{-2})
(\alpha_{-1}\cdot \alpha_{-1})
\,
\,| 0 \rangle $
&14&1&
  & 
 0.00015 & 
 0.00015 
\\
$\psi^{16}$ &  & 
$
(\alpha_{-1}\cdot \alpha_{-2})
(\alpha_{-1}\cdot \alpha_{-2})
\,
\,| 0 \rangle $
&15&1&
  & 
 0.00001 & 
 0.00001 
\\
$\psi^{17}$ &  & 
$
(\alpha_{-1}\cdot \alpha_{-1})
(\alpha_{-1}\cdot \alpha_{-1})
(\alpha_{-1}\cdot \alpha_{-1})
\,
\,| 0 \rangle $
&16&1&
  & 
-0.00000 & 
-0.00000 
\\
$\psi^{18}$ &  & 
$
(\alpha_{-1}\cdot \alpha_{-3})
\,
b_{-1}
c_{-1}
\,| 0 \rangle $
&4&2&
  & 
-0.00263 & 
-0.00280 
\\
$\psi^{19}$ &  & 
$
(\alpha_{-2}\cdot \alpha_{-2})
\,
b_{-1}
c_{-1}
\,| 0 \rangle $
&5&2&
  & 
-0.00116 & 
-0.00121 
\\
$\psi^{20}$ &  & 
$
(\alpha_{-1}\cdot \alpha_{-1})
(\alpha_{-1}\cdot \alpha_{-1})
\,
b_{-1}
c_{-1}
\,| 0 \rangle $
&6&2&
  & 
-0.00008 & 
-0.00009 
\\
$\psi^{21}$ &  & 
$
(\alpha_{-1}\cdot \alpha_{-2})
\,
b_{-1}
c_{-2}
\,| 0 \rangle $
&3&3&
  & 
-0.00013 & 
-0.00015 
\\
$\psi^{22}$ &  & 
$
(\alpha_{-1}\cdot \alpha_{-2})
\,
b_{-2}
c_{-1}
\,| 0 \rangle $
&3&4&
  & 
-0.00006 & 
-0.00007 
\\
$\psi^{23}$ &  & 
$
(\alpha_{-1}\cdot \alpha_{-1})
\,
b_{-1}
c_{-3}
\,| 0 \rangle $
&2&5&
  & 
-0.00336 & 
-0.00351 
\\
$\psi^{24}$ &  & 
$
(\alpha_{-1}\cdot \alpha_{-1})
\,
b_{-2}
c_{-2}
\,| 0 \rangle $
&2&6&
  & 
-0.00283 & 
-0.00295 
\\
$\psi^{25}$ &  & 
$
(\alpha_{-1}\cdot \alpha_{-1})
\,
b_{-3}
c_{-1}
\,| 0 \rangle $
&2&7&
  & 
-0.00112 & 
-0.00117 
\\
$\psi^{26}$ &  & 
$
\,
b_{-1}
c_{-5}
\,| 0 \rangle $
&1&12&
  & 
-0.06003 & 
-0.06094 
\\
$\psi^{27}$ &  & 
$
\,
b_{-2}
c_{-4}
\,| 0 \rangle $
&1&13&
  & 
-0.03750 & 
-0.03816 
\\
$\psi^{28}$ &  & 
$
\,
b_{-3}
c_{-3}
\,| 0 \rangle $
&1&14&
  & 
-0.02283 & 
-0.02291 
\\
$\psi^{29}$ &  & 
$
\,
b_{-4}
c_{-2}
\,| 0 \rangle $
&1&15&
  & 
-0.01875 & 
-0.01908 
\\
$\psi^{30}$ &  & 
$
\,
b_{-5}
c_{-1}
\,| 0 \rangle $
&1&16&
  & 
-0.01201 & 
-0.01219 
\\
$\psi^{31}$ &  & 
$
\,
b_{-2}
b_{-1}
c_{-2}
c_{-1}
\,| 0 \rangle $
&1&17&
  & 
 0.01547 & 
 0.01612 
\\
\hline
\hline
\end{tabular}
\end{center}

\newpage

\section{Table of matter and ghost states  at levels $\leq 6$}

The following tables list the matter and ghost states which contribute
to scalar fields at levels $\leq 6$.  The quadratic terms involving
each state are listed in the last column.
\vspace{0.1in}

\begin{center}
Matter states\\[0.1in]
\begin{tabular}{| r | r | l |}
\hline
\hline
$| \eta_m \rangle $ & state & quadratic terms\\
\hline
\hline
$|\eta_{1}\rangle$ & 
$
\,
\,| 0 \rangle $
 \rule[-0.1cm]{0cm}{0.56cm} & 
$d^{\rm mat}_{1\,1} = $
$-\frac{
1
}{
2^{}
}$
\\[0.1cm]
\hline
$|\eta_{2}\rangle$ & 
$
(\alpha_{-1}\cdot \alpha_{-1})
\,
\,| 0 \rangle $
 \rule[-0.1cm]{0cm}{0.56cm} & 
$d^{\rm mat}_{2\,2} = $
$
2^{}
\cdot
13^{}
$
\\[0.1cm]
\hline
$|\eta_{3}\rangle$ & 
$
(\alpha_{-1}\cdot \alpha_{-2})
\,
\,| 0 \rangle $
 \rule[-0.1cm]{0cm}{0.56cm} & 
$d^{\rm mat}_{3\,3} = $
$
2^{2}
\cdot
13^{}
$
\\[0.1cm]
\hline
$|\eta_{4}\rangle$ & 
$
(\alpha_{-1}\cdot \alpha_{-3})
\,
\,| 0 \rangle $
 \rule[-0.1cm]{0cm}{0.56cm} & 
$d^{\rm mat}_{4\,4} = $
$
3^{2}
\cdot
13^{}
$
\\
$|\eta_{5}\rangle$ & 
$
(\alpha_{-2}\cdot \alpha_{-2})
\,
\,| 0 \rangle $
 \rule[-0.1cm]{0cm}{0.56cm} & 
$d^{\rm mat}_{5\,5} = $
$
2^{3}
\cdot
3^{}
\cdot
13^{}
$
\\
$|\eta_{6}\rangle$ & 
$
(\alpha_{-1}\cdot \alpha_{-1})
(\alpha_{-1}\cdot \alpha_{-1})
\,
\,| 0 \rangle $
 \rule[-0.1cm]{0cm}{0.56cm} & 
$d^{\rm mat}_{6\,6} = $
$
2^{5}
\cdot
3^{}
\cdot
7^{}
\cdot
13^{}
$
\\[0.1cm]
\hline
$|\eta_{10}\rangle$ & 
$
(\alpha_{-1}\cdot \alpha_{-5})
\,
\,| 0 \rangle $
 \rule[-0.1cm]{0cm}{0.56cm} & 
$d^{\rm mat}_{10\,10} = $
$
5^{2}
\cdot
13^{}
$
\\
$|\eta_{11}\rangle$ & 
$
(\alpha_{-2}\cdot \alpha_{-4})
\,
\,| 0 \rangle $
 \rule[-0.1cm]{0cm}{0.56cm} & 
$d^{\rm mat}_{11\,11} = $
$
2^{3}
\cdot
5^{}
\cdot
13^{}
$
\\
$|\eta_{12}\rangle$ & 
$
(\alpha_{-3}\cdot \alpha_{-3})
\,
\,| 0 \rangle $
 \rule[-0.1cm]{0cm}{0.56cm} & 
$d^{\rm mat}_{12\,12} = $
$
2^{}
\cdot
3^{2}
\cdot
5^{}
\cdot
13^{}
$
\\
$|\eta_{13}\rangle$ & 
$
(\alpha_{-1}\cdot \alpha_{-3})
(\alpha_{-1}\cdot \alpha_{-1})
\,
\,| 0 \rangle $
 \rule[-0.1cm]{0cm}{0.56cm} & 
$d^{\rm mat}_{13\,13} = $
$
2^{3}
\cdot
3^{}
\cdot
5^{}
\cdot
7^{}
\cdot
13^{}
$
\\
$|\eta_{14}\rangle$ & 
$
(\alpha_{-2}\cdot \alpha_{-2})
(\alpha_{-1}\cdot \alpha_{-1})
\,
\,| 0 \rangle $
 \rule[-0.1cm]{0cm}{0.56cm} & 
$d^{\rm mat}_{14\,14} = $
$
2^{5}
\cdot
5^{}
\cdot
13^{2}
$
, \,\,\,
$d^{\rm mat}_{14\,15} = $
$
2^{4}
\cdot
5^{}
\cdot
13^{}
$
\\
$|\eta_{15}\rangle$ & 
$
(\alpha_{-1}\cdot \alpha_{-2})
(\alpha_{-1}\cdot \alpha_{-2})
\,
\,| 0 \rangle $
 \rule[-0.1cm]{0cm}{0.56cm} & 
$d^{\rm mat}_{15\,14} = $
$
2^{4}
\cdot
5^{}
\cdot
13^{}
$
, \,\,\,
$d^{\rm mat}_{15\,15} = $
$
2^{3}
\cdot
3^{3}
\cdot
5^{}
\cdot
13^{}
$
\\
$|\eta_{16}\rangle$ & 
$
(\alpha_{-1}\cdot \alpha_{-1})
(\alpha_{-1}\cdot \alpha_{-1})
(\alpha_{-1}\cdot \alpha_{-1})
\,
\,| 0 \rangle $
 \rule[-0.1cm]{0cm}{0.56cm} & 
$d^{\rm mat}_{16\,16} = $
$
2^{7}
\cdot
3^{2}
\cdot
5^{2}
\cdot
7^{}
\cdot
13^{}
$
\\[0.1cm]
\hline
\hline
\end{tabular}
\end{center}

\vspace{0.1in}

\begin{center}
Ghost states\\[0.1in]
\begin{tabular}{| r | r | l |}
\hline
\hline
$| \chi_g \rangle $ & state & quadratic terms\\
\hline
\hline
$|\chi_{1}\rangle$ & 
$
\,
\,| 0 \rangle $
 \rule[-0.1cm]{0cm}{0.56cm} & 
$d^{\rm  gh}_{1\,1} = $
$-\frac{
1
}{
2^{}
}$
\\[0.1cm]
\hline
$|\chi_{2}\rangle$ & 
$
\,
b_{-1}
c_{-1}
\,| 0 \rangle $
 \rule[-0.1cm]{0cm}{0.56cm} & 
$d^{\rm  gh}_{2\,2} = $
$-\frac{
1
}{
2^{}
}$
\\[0.1cm]
\hline
$|\chi_{3}\rangle$ & 
$
\,
b_{-1}
c_{-2}
\,| 0 \rangle $
 \rule[-0.1cm]{0cm}{0.56cm} & 
$d^{\rm  gh}_{3\,4} = $
$
-1^{}
$
\\[0.1cm]
$|\chi_{4}\rangle$ & 
$
\,
b_{-2}
c_{-1}
\,| 0 \rangle $
 \rule[-0.1cm]{0cm}{0.56cm} & 
$d^{\rm  gh}_{4\,3} = $
$
-1^{}
$
\\[0.1cm]
\hline
$|\chi_{5}\rangle$ & 
$
\,
b_{-1}
c_{-3}
\,| 0 \rangle $
 \rule[-0.1cm]{0cm}{0.56cm} & 
$d^{\rm  gh}_{5\,7} = $
$-\frac{
3^{}
}{
2^{}
}$
\\[0.1cm]
$|\chi_{6}\rangle$ & 
$
\,
b_{-2}
c_{-2}
\,| 0 \rangle $
 \rule[-0.1cm]{0cm}{0.56cm} & 
$d^{\rm  gh}_{6\,6} = $
$-\frac{
3^{}
}{
2^{}
}$
\\[0.1cm]
$|\chi_{7}\rangle$ & 
$
\,
b_{-3}
c_{-1}
\,| 0 \rangle $
 \rule[-0.1cm]{0cm}{0.56cm} & 
$d^{\rm  gh}_{7\,5} = $
$-\frac{
3^{}
}{
2^{}
}$
\\[0.1cm]
\hline
$|\chi_{12}\rangle$ & 
$
\,
b_{-1}
c_{-5}
\,| 0 \rangle $
 \rule[-0.1cm]{0cm}{0.56cm} & 
$d^{\rm  gh}_{12\,16} = $
$-\frac{
5^{}
}{
2^{}
}$
\\
$|\chi_{13}\rangle$ & 
$
\,
b_{-2}
c_{-4}
\,| 0 \rangle $
 \rule[-0.1cm]{0cm}{0.56cm} & 
$d^{\rm  gh}_{13\,15} = $
$-\frac{
5^{}
}{
2^{}
}$
\\
$|\chi_{14}\rangle$ & 
$
\,
b_{-3}
c_{-3}
\,| 0 \rangle $
 \rule[-0.1cm]{0cm}{0.56cm} & 
$d^{\rm  gh}_{14\,14} = $
$-\frac{
5^{}
}{
2^{}
}$
\\
$|\chi_{15}\rangle$ & 
$
\,
b_{-4}
c_{-2}
\,| 0 \rangle $
 \rule[-0.1cm]{0cm}{0.56cm} & 
$d^{\rm  gh}_{15\,13} = $
$-\frac{
5^{}
}{
2^{}
}$
\\
$|\chi_{16}\rangle$ & 
$
\,
b_{-5}
c_{-1}
\,| 0 \rangle $
 \rule[-0.1cm]{0cm}{0.56cm} & 
$d^{\rm  gh}_{16\,12} = $
$-\frac{
5^{}
}{
2^{}
}$
\\
$|\chi_{17}\rangle$ & 
$
\,
b_{-2}
b_{-1}
c_{-2}
c_{-1}
\,| 0 \rangle $
 \rule[-0.1cm]{0cm}{0.56cm} & 
$d^{\rm  gh}_{17\,17} = $
$\frac{
5^{}
}{
2^{}
}$
\\[0.1cm]
\hline
\hline
\end{tabular}
\end{center}

\newpage

\section{Cubic interactions  at levels (6, 16)}

In the following pages we give tables of all cubic interactions
between the pure matter and ghost scalar fields listed in Appendix B
which contribute to cubic interactions of total dimension $\leq 16$
between the scalar fields listed in Appendix A.  All cubic
interactions of level less or equal to (6, 16) between the scalar
fields listed in Appendix A can be reproduced using these tables and
(\ref{eq:coefficient-product}).

\vspace{0.1in}
\scriptsize

\begin{center}
{\normalsize Cubic matter field coefficients $t^{{\rm mat}}_{ijk}$}
\vspace{0.1in}

\tabcolsep=0.1mm
\begin{tabular}{||  c |  c || c | c | c | c | c| c||}
\hline
\hline
\multicolumn{2}{||c  ||}{} &\multicolumn{6}{c ||}{i}\\
\hline
{\scriptsize
$\; \,j \; $} &
{\scriptsize$  \;\,  k  \; $} &
{\scriptsize 1} &
{\scriptsize 2} &
{\scriptsize 3} &
{\scriptsize 4} &
{\scriptsize 5}  &
{\scriptsize 6} \\
\hline
\hline
\rule[-0.2cm]{0cm}{0.56cm} 
1 & 1 
 & 
$
1
$
 & 
$-\frac{
2^{}
\cdot
5^{}
\cdot
13^{}
}{
3^{3}
}$
 & 
$
0^{}
$
 & 
$\frac{
2^{6}
\cdot
13^{}
}{
3^{5}
}$
 & 
$\frac{
2^{2}
\cdot
13^{2}
}{
3^{5}
}$
 & 
$\frac{
2^{3}
\cdot
5^{2}
\cdot
7^{}
\cdot
13^{}
}{
3^{6}
}$
\\ \hline
\rule[-0.2cm]{0cm}{0.56cm} 
1 & 2 
 & 
$-\frac{
2^{}
\cdot
5^{}
\cdot
13^{}
}{
3^{3}
}$
 & 
$\frac{
2^{2}
\cdot
7^{}
\cdot
13^{}
\cdot
83^{}
}{
3^{6}
}$
 & 
$\frac{
2^{12}
\cdot
13^{}
\, \sqrt{3}}{
3^{8}
}$
 & 
$-\frac{
2^{7}
\cdot
13^{}
\cdot
73^{}
}{
3^{8}
}$
 & 
$-\frac{
2^{3}
\cdot
13^{}
\cdot
487^{}
}{
3^{9}
}$
 & 
$-\frac{
2^{4}
\cdot
5^{}
\cdot
7^{}
\cdot
13^{}
\cdot
31^{}
}{
3^{6}
}$
\\ \hline
\rule[-0.2cm]{0cm}{0.56cm} 
1 & 3 
 & 
$
0^{}
$
 & 
$-\frac{
2^{12}
\cdot
13^{}
\, \sqrt{3}}{
3^{8}
}$
 & 
$-\frac{
2^{15}
\cdot
13^{}
}{
3^{9}
}$
 & 
$-\frac{
2^{12}
\cdot
7^{}
\cdot
13^{}
\, \sqrt{3}}{
3^{10}
}$
 & 
$-\frac{
2^{16}
\cdot
13^{}
\, \sqrt{3}}{
3^{10}
}$
 & 
$\frac{
2^{15}
\cdot
5^{}
\cdot
7^{}
\cdot
13^{}
\, \sqrt{3}}{
3^{11}
}$
\\ \hline
\rule[-0.2cm]{0cm}{0.56cm} 
1 & 4 
 & 
$\frac{
2^{6}
\cdot
13^{}
}{
3^{5}
}$
 & 
$-\frac{
2^{7}
\cdot
13^{}
\cdot
73^{}
}{
3^{8}
}$
 & 
$\frac{
2^{12}
\cdot
7^{}
\cdot
13^{}
\, \sqrt{3}}{
3^{10}
}$
 & 
$\frac{
2^{10}
\cdot
13^{}
\cdot
479^{}
}{
3^{11}
}$
 & 
$\frac{
2^{8}
\cdot
13^{}
\cdot
163^{}
}{
3^{9}
}$
 & 
$\frac{
2^{9}
\cdot
5^{}
\cdot
7^{}
\cdot
13^{}
}{
3^{7}
}$
\\
\hline
\hline
\end{tabular}
\end{center}

\newpage

\begin{center}
{\normalsize Table of $t^{{\rm mat}}_{ijk}$  (continued)}
\vspace{0.4in}

\tabcolsep=0.1mm
\begin{tabular}{||  c |  c || c | c | c | c | c| c||}
\hline
\hline
\multicolumn{2}{||c  ||}{} &\multicolumn{6}{c ||}{i}\\
\hline
{\scriptsize
$\; \,j \; $} &
{\scriptsize$  \;\,  k  \; $} &
{\scriptsize 1} &
{\scriptsize 2} &
{\scriptsize 3} &
{\scriptsize 4} &
{\scriptsize 5}  &
{\scriptsize 6} \\
\hline
\hline
\rule[-0.2cm]{0cm}{0.56cm} 
3 & 10 
 & 
$\frac{
2^{12} % \cdot
5^{} \cdot
13^{} \cdot
43^{}
\sqrt{3}}{
3^{14}
}$
 & 
$-\frac{
2^{13} % \cdot
5^{} \cdot
13^{} \cdot
7307^{}
\sqrt{3}}{
3^{17}
}$
 & 
$\frac{
2^{19} % \cdot
5^{} \cdot
13^{2} % \cdot
17^{}
}{
3^{17}
}$
 & 
$-\frac{
2^{21} % \cdot
5^{3} % \cdot
13^{}
\sqrt{3}}{
3^{18}
}$
 & 
$-\frac{
2^{14} % \cdot
5^{} \cdot
13^{} \cdot
149^{} \cdot
233^{}
\sqrt{3}}{
3^{19}
}$
 & 
$\frac{
2^{15} % \cdot
5^{} \cdot
7^{2} % \cdot
13^{} \cdot
31^{} \cdot
101^{}
\sqrt{3}}{
3^{19}
}$
\\ \hline
\rule[-0.2cm]{0cm}{0.56cm} 
3 & 11 
 & 
$\frac{
2^{18} % \cdot
5^{} \cdot
13^{}
\sqrt{3}}{
3^{14}
}$
 & 
$-\frac{
2^{19} % \cdot
13^{} \cdot
773^{}
\sqrt{3}}{
3^{17}
}$
 & 
$\frac{
2^{25} % \cdot
13^{} \cdot
17^{}
}{
3^{17}
}$
 & 
$\frac{
2^{24} % \cdot
13^{}
\sqrt{3}}{
3^{16}
}$
 & 
$-\frac{
2^{20} % \cdot
5^{} \cdot
13^{} \cdot
599^{}
\sqrt{3}}{
3^{19}
}$
 & 
$\frac{
2^{21} % \cdot
7^{} \cdot
13^{} \cdot
1523^{}
\sqrt{3}}{
3^{19}
}$
\\ \hline
\rule[-0.2cm]{0cm}{0.56cm} 
3 & 12 
 & 
$\frac{
2^{12} % \cdot
5^{} \cdot
13^{}
\sqrt{3}}{
3^{11}
}$
 & 
$-\frac{
2^{14} % \cdot
13^{} \cdot
6883^{}
\sqrt{3}}{
3^{16}
}$
 & 
$\frac{
2^{16} % \cdot
13^{} \cdot
8513^{}
}{
3^{17}
}$
 & 
$-\frac{
2^{13} % \cdot
13^{} \cdot
144847^{}
\sqrt{3}}{
3^{18}
}$
 & 
$-\frac{
2^{14} % \cdot
13^{} \cdot
50707^{}
\sqrt{3}}{
3^{18}
}$
 & 
$\frac{
2^{15} % \cdot
7^{} \cdot
13^{} \cdot
43^{} \cdot
2897^{}
\sqrt{3}}{
3^{19}
}$
\\
\hline
\hline
\end{tabular}
\end{center}

\newpage

\begin{center}
{\normalsize Table of $t^{{\rm mat}}_{ijk}$  (continued)}
\vspace{0.3in}

\tabcolsep=0.1mm
\begin{tabular}{||  c |  c || c | c | c | c | c||}
\hline
\hline
\multicolumn{2}{||c  ||}{} &\multicolumn{5}{c ||}{i}\\
\hline
{\scriptsize
$\; \,j \; $} &
{\scriptsize$  \;\,  k  \; $} &
{\scriptsize 1} &
{\scriptsize 2} &
{\scriptsize 4} &
{\scriptsize 5}  &
{\scriptsize 6} \\
\hline
\hline
\rule[-0.2cm]{0cm}{0.56cm} 
10 & 10 
 & 
$\frac{
2^{10}
\cdot
5^{}
\cdot
13^{}
\cdot
109^{}
\cdot
839^{}
}{
3^{17}
}$
 & 
$-\frac{
2^{11}
\cdot
5^{}
\cdot
13^{}
\cdot
251^{}
\cdot
59107^{}
}{
3^{21}
}$
 & 
$\frac{
2^{17}
\cdot
5^{2}
\cdot
7^{}
\cdot
13^{}
\cdot
17033^{}
}{
3^{22}
}$
 & 
$\frac{
2^{12}
\cdot
5^{}
\cdot
13^{}
\cdot
16686323^{}
}{
3^{22}
}$
 & 
$\frac{
2^{13}
\cdot
5^{2}
\cdot
7^{}
\cdot
13^{}
\cdot
31^{}
\cdot
42743^{}
}{
3^{22}
}$
\\ \hline
\rule[-0.2cm]{0cm}{0.56cm} 
10 & 11 
 & 
$\frac{
2^{16}
\cdot
5^{}
\cdot
13^{}
\cdot
29^{}
}{
3^{14}
}$
 & 
$-\frac{
2^{17}
\cdot
5^{}
\cdot
13^{}
\cdot
168013^{}
}{
3^{21}
}$
 & 
$\frac{
2^{27}
\cdot
5^{}
\cdot
13^{}
\cdot
599^{}
}{
3^{23}
}$
 & 
$\frac{
2^{18}
\cdot
5^{}
\cdot
13^{}
\cdot
768437^{}
}{
3^{23}
}$
 & 
$\frac{
2^{19}
\cdot
5^{}
\cdot
7^{}
\cdot
13^{}
\cdot
31^{}
\cdot
28927^{}
}{
3^{24}
}$
\\ \hline
\rule[-0.2cm]{0cm}{0.56cm} 
10 & 12 
 & 
$\frac{
2^{7}
\cdot
5^{}
\cdot
7^{2}
\cdot
13^{}
\cdot
47^{}
\cdot
67^{}
}{
3^{17}
}$
 & 
$-\frac{
2^{8}
\cdot
5^{}
\cdot
13^{}
\cdot
11307557^{}
}{
3^{20}
}$
 & 
$\frac{
2^{11}
\cdot
5^{2}
\cdot
13^{}
\cdot
773063^{}
}{
3^{21}
}$
 & 
$-\frac{
2^{9}
\cdot
5^{}
\cdot
13^{}
\cdot
709^{}
\cdot
37567^{}
}{
3^{22}
}$
 & 
$\frac{
2^{10}
\cdot
5^{}
\cdot
7^{}
\cdot
13^{}
\cdot
47^{}
\cdot
149143^{}
}{
3^{21}
}$
\\
\hline
\hline
\end{tabular}
\end{center}

\newpage

\begin{center}
{\normalsize Cubic ghost field coefficients $t^{{\rm gh}}_{ijk}$}
\vspace{0.3in}

\tabcolsep=0.1mm
\begin{tabular}{||  c |  c || c | c | c | c | c| c | c ||}
\hline
\hline
\multicolumn{2}{||c  ||}{} &\multicolumn{7}{c ||}{i}\\
\hline
{\scriptsize
$\; \,j \; $} &
{\scriptsize$  \;\,  k  \; $} &
{\scriptsize 1} &
{\scriptsize 2} &
{\scriptsize 3} &
{\scriptsize 4} &
{\scriptsize 5}  &
{\scriptsize 6}  &
{\scriptsize 7} \\
\hline
\hline
\rule[-0.2cm]{0cm}{0.56cm} 
1 & 1 
 & 
$
1
$
 & 
$-\frac{
11^{}
}{
3^{3}
}$
 & 
$
0^{}
$
 & 
$
0^{}
$
 & 
$\frac{
2^{4}
\cdot
5^{}
}{
3^{6}
}$
 & 
$\frac{
19^{}
}{
3^{5}
}$
 & 
$\frac{
2^{4}
\cdot
5^{}
}{
3^{5}
}$
\\ \hline
\rule[-0.2cm]{0cm}{0.56cm} 
1 & 2 
 & 
$-\frac{
11^{}
}{
3^{3}
}$
 & 
$\frac{
19^{}
}{
3^{5}
}$
 & 
$-\frac{
2^{6}
\cdot
5^{}
\, \sqrt{3}}{
3^{8}
}$
 & 
$-\frac{
2^{7}
\cdot
5^{}
\, \sqrt{3}}{
3^{8}
}$
 & 
$-\frac{
2^{4}
\cdot
5^{2}
}{
3^{8}
}$
 & 
$-\frac{
43^{}
\cdot
89^{}
}{
3^{9}
}$
 & 
$-\frac{
2^{4}
\cdot
5^{2}
}{
3^{7}
}$
\\ \hline
\rule[-0.2cm]{0cm}{0.56cm} 
1 & 3 
 & 
$
0^{}
$
 & 
$\frac{
2^{6}
\cdot
5^{}
\, \sqrt{3}}{
3^{8}
}$
 & 
$\frac{
2^{6}
\cdot
5^{2}
}{
3^{9}
}$
 & 
$\frac{
2^{7}
\cdot
7^{}
}{
3^{8}
}$
 & 
$\frac{
2^{6}
\cdot
5^{2}
\, \sqrt{3}}{
3^{11}
}$
 & 
$\frac{
2^{7}
\cdot
5^{}
\cdot
7^{}
\, \sqrt{3}}{
3^{10}
}$
 & 
$\frac{
2^{6}
\cdot
19^{}
\, \sqrt{3}}{
3^{9}
}$
\\ \hline
\rule[-0.2cm]{0cm}{0.56cm} 
1 & 4 
 & 
$
0^{}
$
 & 
$\frac{
2^{7}
\cdot
5^{}
\, \sqrt{3}}{
3^{8}
}$
 & 
$\frac{
2^{7}
\cdot
7^{}
}{
3^{8}
}$
 & 
$\frac{
2^{8}
\cdot
5^{2}
}{
3^{9}
}$
 & 
$\frac{
2^{7}
\cdot
19^{}
\, \sqrt{3}}{
3^{10}
}$
 & 
$\frac{
2^{8}
\cdot
5^{}
\cdot
7^{}
\, \sqrt{3}}{
3^{10}
}$
 & 
$\frac{
2^{7}
\cdot
5^{2}
\, \sqrt{3}}{
3^{10}
}$
\\
\hline
\hline
\end{tabular}
\end{center}

\newpage

\begin{center}
{\normalsize Table of $t^{{\rm gh}}_{ijk}$  (continued)}
\vspace{0.3in}

\tabcolsep=0.1mm
\begin{tabular}{||  c |  c || c | c | c | c | c| c | c ||}
\hline
\hline
\multicolumn{2}{||c  ||}{} &\multicolumn{7}{c ||}{i}\\
\hline
{\scriptsize
$\; \,j \; $} &
{\scriptsize$  \;\,  k  \; $} &
{\scriptsize 1} &
{\scriptsize 2} &
{\scriptsize 3} &
{\scriptsize 4} &
{\scriptsize 5}  &
{\scriptsize 6} &
{\scriptsize 7} \\
\hline
\hline
\rule[-0.2cm]{0cm}{0.56cm} 
3 & 12 
 & 
$-\frac{
2^{6}
\cdot
5^{}
\cdot
71^{}
\, \sqrt{3}}{
3^{14}
}$
 & 
$\frac{
2^{6}
\cdot
5^{}
\cdot
71^{}
\, \sqrt{3}}{
3^{14}
}$
 & 
$
0^{}
$
 & 
$-\frac{
2^{13}
\cdot
89^{}
}{
3^{16}
}$
 & 
$
0^{}
$
 & 
$\frac{
2^{6}
\cdot
5^{}
\cdot
5641^{}
\, \sqrt{3}}{
3^{18}
}$
 & 
$\frac{
2^{14}
\cdot
373^{}
\, \sqrt{3}}{
3^{19}
}$
\\ \hline
\rule[-0.2cm]{0cm}{0.56cm} 
3 & 13 
 & 
$-\frac{
2^{7}
\cdot
7^{}
\cdot
13^{}
\, \sqrt{3}}{
3^{12}
}$
 & 
$\frac{
2^{7}
\cdot
43^{}
\cdot
227^{}
\, \sqrt{3}}{
3^{17}
}$
 & 
$-\frac{
2^{15}
\cdot
5^{}
\cdot
59^{}
}{
3^{18}
}$
 & 
$\frac{
2^{14}
\cdot
5^{}
\cdot
7^{}
}{
3^{16}
}$
 & 
$-\frac{
2^{13}
\cdot
5^{}
\cdot
349^{}
\, \sqrt{3}}{
3^{20}
}$
 & 
$\frac{
2^{7}
\cdot
7^{}
\cdot
17^{}
\cdot
331^{}
\, \sqrt{3}}{
3^{18}
}$
 & 
$-\frac{
2^{13}
\cdot
5^{2}
\cdot
71^{}
\, \sqrt{3}}{
3^{18}
}$
\\ \hline
\rule[-0.2cm]{0cm}{0.56cm} 
3 & 14 
 & 
$\frac{
2^{6}
\cdot
5^{}
\cdot
19^{}
\, \sqrt{3}}{
3^{12}
}$
 & 
$\frac{
2^{7}
\cdot
5^{}
\cdot
17^{}
\cdot
29^{}
\, \sqrt{3}}{
3^{17}
}$
 & 
$-\frac{
2^{6}
\cdot
5^{3}
\cdot
967^{}
}{
3^{18}
}$
 & 
$\frac{
2^{7}
\cdot
11^{}
\cdot
1409^{}
}{
3^{17}
}$
 & 
$-\frac{
2^{6}
\cdot
5^{2}
\cdot
863^{}
\, \sqrt{3}}{
3^{19}
}$
 & 
$-\frac{
2^{6}
\cdot
5^{}
\cdot
197^{}
\, \sqrt{3}}{
3^{19}
}$
 & 
$\frac{
2^{6}
\cdot
19^{}
\cdot
11491^{}
\, \sqrt{3}}{
3^{18}
}$
\\
\hline
\hline
\end{tabular}
\end{center}

\begin{center}
{\normalsize Table of $t^{{\rm gh}}_{ijk}$  (continued)}
\vspace{0.3in}

\tabcolsep=0.1mm
\begin{tabular}{||  c |  c || c | c | c | c | c ||}
\hline
\hline
\multicolumn{2}{||c  ||}{} &\multicolumn{5}{c ||}{i}\\
\hline
{\scriptsize
$\; \,j \; $} &
{\scriptsize$  \;\,  k  \; $} &
{\scriptsize 1} &
{\scriptsize 2} &
{\scriptsize 5} &
{\scriptsize 6}  &
{\scriptsize 7} \\
\hline
\hline
\rule[-0.2cm]{0cm}{0.56cm} 
12 & 12 
 & 
$\frac{
2^{6}
\cdot
71^{2}
}{
3^{17}
}$
 & 
$-\frac{
2^{6}
\cdot
71^{2}
}{
3^{17}
}$
 & 
$
0^{}
$
 & 
$-\frac{
2^{6}
\cdot
71^{}
\cdot
5531^{}
}{
3^{22}
}$
 & 
$\frac{
2^{20}
\cdot
71^{}
}{
3^{22}
}$
\\ \hline
\rule[-0.2cm]{0cm}{0.56cm} 
12 & 13 
 & 
$-\frac{
2^{7}
\cdot
7993^{}
}{
3^{18}
}$
 & 
$\frac{
2^{7}
\cdot
7723^{}
}{
3^{19}
}$
 & 
$-\frac{
2^{14}
\cdot
11^{2}
\cdot
31^{}
}{
3^{23}
}$
 & 
$-\frac{
2^{7}
\cdot
189187^{}
}{
3^{21}
}$
 & 
$\frac{
2^{14}
\cdot
5^{2}
\cdot
563^{}
}{
3^{23}
}$
\\ \hline
\rule[-0.2cm]{0cm}{0.56cm} 
12 & 14 
 & 
$\frac{
2^{4}
\cdot
11^{}
\cdot
19^{}
\cdot
521^{}
}{
3^{18}
}$
 & 
$-\frac{
2^{4}
\cdot
5^{}
\cdot
106921^{}
}{
3^{20}
}$
 & 
$\frac{
2^{6}
\cdot
5^{}
\cdot
67^{}
\cdot
10343^{}
}{
3^{23}
}$
 & 
$\frac{
2^{4}
\cdot
25908643^{}
}{
3^{23}
}$
 & 
$\frac{
2^{6}
\cdot
7^{2}
\cdot
13^{}
\cdot
17^{}
\cdot
109^{}
}{
3^{22}
}$
\\ \hline
\rule[-0.2cm]{0cm}{0.56cm} 
12 & 15 
 & 
$-\frac{
2^{8}
\cdot
5^{}
\cdot
277^{}
}{
3^{14}
}$
 & 
$\frac{
2^{8}
\cdot
5^{}
\cdot
409831^{}
}{
3^{21}
}$
 & 
$-\frac{
2^{15}
\cdot
5^{2}
\cdot
11^{}
\cdot
23^{}
}{
3^{24}
}$
 & 
$-\frac{
2^{8}
\cdot
5^{}
\cdot
41^{}
\cdot
74887^{}
}{
3^{23}
}$
 & 
$-\frac{
2^{15}
\cdot
1213^{}
}{
3^{21}
}$
\\ \hline
\rule[-0.2cm]{0cm}{0.56cm} 
12 & 16 
 & 
$-\frac{
2^{6}
\cdot
67^{}
\cdot
11083^{}
}{
3^{18}
}$
 & 
$\frac{
2^{6}
\cdot
7^{2}
\cdot
53^{2}
}{
3^{20}
}$
 & 
$-\frac{
2^{22}
\cdot
5^{}
\cdot
7^{}
\cdot
13^{}
}{
3^{23}
}$
 & 
$\frac{
2^{6}
\cdot
13^{}
\cdot
229^{}
\cdot
29453^{}
}{
3^{23}
}$
 & 
$-\frac{
2^{22}
\cdot
5^{}
\cdot
7^{}
\cdot
13^{}
}{
3^{22}
}$
\\
\hline
\hline
\end{tabular}
\end{center}

\normalsize
\newpage

\bibliographystyle{plain}
%\bibliography{papers}

\end{document}